\documentclass{osa-article}

\journal{osajournal}

\articletype{Research Article}

\usepackage{algorithm,algorithmic,minibox,multirow,hhline}
\usepackage{float,verbatim,amsmath,amsfonts,bm}
\usepackage{xcolor,makecell}

\begin{document}
	
\title{The CRLB, variance, bias and Maximum likelihood in ptychography with Poisson noise}
	
	\author{Xukang Wei,\authormark{1,*} H. Paul Urbach\authormark{1} and W. M. J. Coene\authormark{1,2}}
	
	\address{\authormark{1}Optics Research Group, Imaging Physics Department, Delft University of Technology, The Netherlands
	\authormark{2}ASML Netherlands B.V, The Netherlands}
	
	\email{\authormark{*}x.wei-2@tudelft.nl} 
	
	
	
\begin{abstract*}
	We investigate the performance of ptychography with noisy data by analyzing the Cram\'{e}r Rao Lower Bound. The lower bound of ptychography is derived and numerically computed for both plane wave and structured illumination. The influence of Poisson noise on the ptychography reconstruction is discussed. The computation result shows that, if the estimator is unbiased, the minimum variance for Poisson noise is mostly determined by the illumination power and the transmission function of the object. Monte Carlo analysis is conducted to validate our calculation results for different photon numbers. The performance of the maximum likelihood method and the approach of amplitude-based cost function minimization is studied in the Monte Carlo analysis also.
\end{abstract*}
	
\section{Introduction}

Ptychography\cite{Hoppe1969,Rodenburg1992a,Chapman1996,Faulkner2004,Rodenburg2004,Guizar-Sicairos2008} is a scanning coherent diffraction imaging method for reconstructing a complex valued object function from intensity measurements recorded in the Fraunhofer or Fresnel diffraction region. In ptychography the object is partially illuminated multiple times so that the entire object is covered and adjacent illuminations partially overlap \cite{Silva2015}. The technique is found very suitable for EUV \cite{Seaberg2014,Odstrcil2015} and X-ray imaging applications \cite{Rodenburg2007,Thibault2008,Chapman2010,Pfeiffer2017} due to its high fidelity and its minimum requirement on optical imaging elements. Moreover, abundant studies show that ptychography is able to provide a wide field-of-view and retrieve the illumination probe also \cite{Thibault2009,Maiden2009}. During the last two decades, ptychography has been successfully demonstrated with X-ray radiation sources \cite{Thibault2008,Holler2017,Gardner2017}, electron beams \cite{Jiang2018} and visible light sources \cite{Maiden2017}. More recently, many extensions of ptychography have been proposed, including Fourier ptychography\cite{Zheng2013,Yeh2015,Zhang2017}, spatially partial coherent ptychography\cite{Thibault13,Burdet2015,Zhong2016}, broadband ptychography\cite{Batey2014,Wei2019}, 3D ptychography\cite{Maiden2012,Gilles2018,Kahnt2019}, on-the-fly scanning ptychograhy\cite{Pelz2014,Deng2015} and interference probe ptychography\cite{Flaes2018}.

For retrieving the object from a ptychographic data set, the key is to find a solution which fulfills both the ptychographic illumination condition in real space and the corresponding measured diffraction intensities in reciprocal space. A commonly used approach for solving the problem is the ptychography iterative engine\cite{Rodenburg2004,Maiden2009}, which can be derived by sequentially minimizing the distance between the estimated amplitude of the diffracted wavefield and the measurements\cite{Guizar-Sicairos2008}. Another popular choice is the difference map algorithm, which can be formulated in terms of finding the intersection of two constraint sets\cite{Elser2003,Thibault2009}. Based on the augmented Lagrangian methods for solving the conventional constrained optimization problems, several interesting ptychographic algorithm have been developed during the past ten years\cite{Wen2012,Marchesini2013,Horstmeyer2015,Pham2019}.

However, obtaining an unique reconstruction and a reconstruction with minimum defect in ptychography is considered difficult and there is still room for improvement. On the one hand, ambiguities due to a constant scaling factor, a global phase shift and raster grid pathology, occur in particular when the probe is unknown\cite{Fannjiang2019}. Although many algorithms have been presented to enhance the robustness of ptychography \cite{Wen2012,Horstmeyer2015,Maiden2017,Pham2019}, a good starting point and proper parameter settings (e.g. update step size, regularization factor, etc.) are needed in general. Furthermore, noise in the measurements of the diffracted intensity cause inaccuracies in the reconstructions\cite{Thibault2012,Godard2012,Chang2019}. To prevent the effect caused by the saturation of the detector, dark-field and near-field ptychography have been introduced\cite{Suzuki2015,Stockmar2013}. Moreover, it was shown that adaptive step size strategies are able to improve the performance of ptychography in the presence of noise\cite{Zuo2016,Maiden2017}. In general, the most powerful and robust de-noising methods are based on the maximum likelihood principle\cite{Godard2012,Thibault2012,Yeh2015,Odstrcil2018,Chang2019}. The likelihood function used in the maximum likelihood method depends on the noise model. Common choices for the noise model in ptychography are Poisson noise, Gaussian noise and the mixed Poisson-Gaussian model. It has been demonstrated \cite{Godard2012,Thibault2012,Zhang2017,Konijnenberg2018a} that, by using the variance stabilization transform given by Bartlett\cite{Bartlett1936} and Anscombe \cite{Anscombe1948}, one can approximate the maximum likelihood method of Poisson noise by the amplitude-based cost minimization algorithm. Therefore both the approach of maximum likelihood and the amplitude-based cost minimization algorithm can be used as a refinement method in ptychography with noisy data.

In this paper our work contains two parts. In the first part we investigate the Cram\'{e}r Rao Lower Bound (CRLB) for the variance of any unbiased estimator in ptychography\cite{Kay2009,Cederquist1987,Fienup1993}. To the best of our knowledge, this is the first investigation of the CRLB in ptychography. We study the lower bound for Poisson distributed photon counting noise, which is the most dominant source of noise which occurs even under the best experimental conditions\cite{Godard2012,Thibault2012}. In Section 2, we briefly discuss ptychography, Poisson photon counting noise and the maximum likelihood method. We compute the Fisher information matrix of ptychography with Poisson noise and introduce the CRLB. In Section 3, the CRLB is numerical computed and the influence of illumination and of the object is discussed in detail. To validate the obtained CRLB, Monte Carlo analysis is implemented in Section 4. 

For the second part of this paper, the performance of the maximum likelihood method and the approach of amplitude-based cost function minimization are also compared using Monte Carlo simulations. Details of the implementation of the algorithms can be found in Appendix. We investigate the statistical property of the algorithms for various photon counts in Section 4. The paper is concluded with a summary and outlook in the last section.

\section{Theory}

\subsection{Ptychography, Poisson noise, and maximum likelihood method}
The goal of ptychography is to reconstruct a complex-valued object $O$ from a set of diffraction intensity patterns which are recorded in the Fraunhofer or Fresnel region. Let $\textbf{r}$ and $\textbf{r}'$ be 2D coordinates in the object plane and the detector plane, respectively. The exit wave immediately behind the object is denoted by $\varPsi(\textbf{r})$ and the measured diffraction intensity measurement $I(\textbf{r}')$. According to the thin object model, the exit wave $\varPsi(\textbf{r})$ for an illumination with a probe function $P(\textbf{r})$ which is centered on position $\textbf{R}_{m}$ is given by
\begin{align}
\varPsi_{m}(\textbf{r})\,&=\,P(\textbf{r}-\textbf{R}_{m})\cdot O(\textbf{r})\nonumber\\
&=\,P_{m}(\textbf{r})\cdot O(\textbf{r}),\label{eq.4}
\end{align} 
where the object $O(\textbf{r})$ can be decomposed to two real valued functions $A(\textbf{r})$ and $\phi(\textbf{r})$:
\begin{flalign}
O(\textbf{r})\,=\, A(\textbf{r})\cdot e^{\text{i}\phi(\textbf{r})}
\label{eq.28},
\end{flalign}
where $A$ is the object's local transmission function and $\phi$ stands for the phase of the exit wave immediately behind the object.
The probe function is assumed to have a support with circular boundary:
\begin{align}
P(\textbf{r})\,=\,\left\lbrace 
\begin{tabular}{ll}
$P(\textbf{r}),$ & $\left|\textbf{r}\right|\leq r_{0},$\\
$0,$ & $\left|\textbf{r}\right|>r_{0}.$
\end{tabular}
\right.\label{eq.5}
\end{align}

For a detector located at distance $z$ in the far field, the diffraction intensity pattern $I(\textbf{r}')$ for the $m$th illumination is\cite{goodman2005}:
\begin{flalign}
I_{m}(\textbf{r}')\,&=\,\left|\mathcal{F}\left(\varPsi_{m}\right)\left(\frac{\textbf{r}'}{\lambda z}\right)\right|^{2}\nonumber\\
&=\,\left|\sum_{\textbf{r}} \varPsi_{m}(\textbf{r})\cdot\exp\left(-\text{i}\frac{2\pi}{\lambda z}\textbf{r}\cdot\textbf{r}'\right)\right|^{2},\label{eq.6}
\end{flalign}
where $\mathcal{F}$ is the discrete Fourier transform operator. 

The task of ptychography is to find an object function which takes account of the \textit{a priori} knowledge, while a cost function $\mathcal{E}$ is minimized. In our case the \textit{a priori} knowledge is the exact information of the probe function for each relative position $\textbf{R}_{m}$, 
The cost function $\mathcal{E}$ is defined as the $l_{2}$-distance between the modulus of the far field diffraction pattern $\mathcal{F}\left(\varPsi_{m}\right)(\bm{\upxi})$ and the squared root of the measured intensity $I_{m}^{\text{measure}}(\bm{\upxi})$:
\begin{align}
\mathcal{E}\,&=\,\sum_{{m}}\sum_{\bm{\upxi}}\left[\sqrt{I_{m}^{\text{measure}}(\bm{\upxi})}-\left|\mathcal{F}\left(\varPsi_{m}\right)(\bm{\upxi})\right|\right]^{2},\label{eq.8}
\end{align}
where $\bm{\upxi}=\frac{\textbf{r}'}{\lambda z}$ is the spatial spectrum coordinate.

From $I_{m}^{\text{measure}}$, one can estimate the number of detected photons:
\begin{flalign}
n_{m}(\bm{\upxi})\,=\,\frac{I_{m}^{\text{measure}}(\bm{\upxi})}{\hbar\omega},\ \qquad\text{where}\quad\omega\,=\,\frac{2\pi c}{\lambda}.\label{eq.7}
\end{flalign} 
Among various of noise models, we consider Poisson noise. The probability distribution of detecting $n_{m}(\bm{\upxi})$ photons by the detector at every $\bm{\upxi}$ for all $m$th measurements are given by:
\begin{align}
\mathcal{P}_{P}\,&=\,\prod_{{m}}\prod_{\bm{\upxi}}\dfrac{N_{m}(\bm{\upxi})^{n_{m}(\bm{\upxi})}}{n_{m}(\bm{\upxi})!}e^{-N_{m}(\bm{\upxi})},\label{eq.9}
\end{align}
where the cumulative product is over both the 2-D coordinate $\bm{\upxi}$ and the probe position $\textbf{R}_{m}$. The negative log-likelihood functional is defined by:
\begin{align}
\mathcal{L}_{P}\,&=\,-\ln\mathcal{P}_{P}\nonumber\\
&=\,-\sum_{{m}}\sum_{\bm{\upxi}}\left[ n_{m}(\bm{\upxi})\ln N_{m}(\bm{\upxi})-N_{m}(\bm{\upxi})-\ln n_{m}(\bm{\upxi})!\right].\label{eq.10}
\end{align}
The average number of photons $N_{m}(\bm{\upxi})$ depends on the object function $O(\textbf{r})$ through Eq. (\ref{eq.6}) and Eq. (\ref{eq.7}). To find the object function for which the negative log-likelihood functional is maximum, the derivative of $\mathcal{L}_{P}$ with respect to $O$ is set equal to zero. Hence, for any small perturbation $\delta O$ of the object function there should hold: 
\begin{align}
\delta\mathcal{L}_{P}(\delta O)\,&=\,-\sum_{{m}}\sum_{\bm{\upxi}}\left(\frac{n_{m}(\bm{\upxi})}{N_{m}(\bm{\upxi})}-1\right)\delta N_{m}\left(\delta O\right)\nonumber\\
&=\,-\frac{1}{\hbar\omega}\sum_{{m}}\sum_{\bm{\upxi}}\left(\frac{n_{m}(\bm{\upxi})}{N_{m}(\bm{\upxi})}-1\right)\delta I_{m}\left(\delta O\right)\nonumber\\
&=\,-\frac{2}{\hbar\omega}\sum_{{m}}\sum_{\bm{\upxi}}\left(\frac{n_{m}(\bm{\upxi})}{N_{m}(\bm{\upxi})}-1\right)\Re\left[\mathcal{F}\left( P_{m}O \right)(\bm{\upxi})\mathcal{F}\left( P_{m} \delta O \right)(\bm{\upxi})^{*}\right]\nonumber\\
&=\,-\frac{2}{\hbar\omega}\sum_{{m}}\sum_{\textbf{r}}\Re\left\lbrace\mathcal{F}^{-1}\left[\left(\frac{n_{m}(\bm{\upxi})}{N_{m}(\bm{\upxi})}-1\right)\mathcal{F}\left(P_{m}O\right) (\bm{\upxi})\right] P^{*}_{m}(\textbf{r}) \delta O ^{*}(\textbf{r})\right\rbrace\nonumber\\
&=\,0,\label{eq.11}
\end{align}
where Parseval's theorem was used. $\Re$ denotes the real part and $\mathcal{F}^{-1}$ the inverse Fourier transform. The local perturbation of the value of $O$ on a discretized grid $\textbf{r}_{i}$ is written as:
\begin{flalign}
\delta O(\textbf{r})\,=\,\sum_{\textbf{r}_{i}}\delta O(\textbf{r}_{i})\delta(\textbf{r}-\textbf{r}_{i})\,=\,\sum_{\textbf{r}_{i}}
\begin{bmatrix}
\delta A(\textbf{r}_{i})\\
\text{i}A(\textbf{r}_{i})\delta \phi(\textbf{r}_{i})
\end{bmatrix}e^{\text{i}\phi(\textbf{r}_{i})}\delta(\textbf{r}-\textbf{r}_{i}).\label{eq.23}
\end{flalign}
The solution of Eq. (\ref{eq.11}) can be found by the method of steepest descent\cite{Murray1982,Odstrcil2018,Wei2019}:
\begin{flalign}
\left\lbrace 
\begin{aligned}
A_{k+1}(\textbf{r})\,&=\,A_{k}(\textbf{r})+\alpha_{A}\sum_{{m}}\Re\left\lbrace P^{*}_{m}e^{-\text{i}\phi_{k}} \mathcal{F}^{-1}\left[\left(\frac{n_{m}}{N_{m}}-1\right)\mathcal{F}\left(P_{m}O_{k}\right)\right]\right\rbrace(\textbf{r}),\\
\phi_{k+1}(\textbf{r})\,&=\,\phi_{k}(\textbf{r})+\alpha_{\phi}\sum_{{m}}\Im\left\lbrace P^{*}_{m}A_{k} e^{-\text{i}\phi_{k}} \mathcal{F}^{-1}\left[\left(\frac{n_{m}}{N_{m}}-1\right)\mathcal{F}\left(P_{m}O_{k}\right)\right]\right\rbrace(\textbf{r}),
\end{aligned}
\right.\label{eq.12}
\end{flalign}
where $k$ is the iteration number, and $\alpha_{A}$ and $\alpha_{\phi}$ are the step-sizes, which are normally chosen to be a constant, i.e. they are independent on the iteration number. $\Im$ denotes the imaginary part. Alternatively, projection based method or conjugate gradient method can be applied to achieve maximum likelihood\cite{Thibault2012}. 

\subsection{The CRLB and the Fisher matrix}
In estimation theory, the CRLB gives a lower bound on the variance of any unbiased estimator for the parameter which must be estimated. The estimators that can reach the lower bound are called the minimum variance unbiased estimators. Minimum variance unbiased estimators are often not available \cite{Kay2009,Bouchet2020}.

We recall the definition of the CRLB, using the notation as in\cite{Kay2009}. Suppose we wish to retrieve a real valued vector parameter $\mathbf{\Theta}=[\theta_{1},\theta_{2},\cdots]^{T}$ from a set of measurements $\mathbf{X}=[X_{1},X_{2},\cdots]^{T}$. There are infinite number of possible outcomes $\mathbf{X}_{1},\mathbf{X}_{2},\cdots,\mathbf{X}_{s},\cdots$ occurring with probabilities $\mathcal{P}_{1},\mathcal{P}_{2},\cdots,\mathcal{P}_{s},\cdots$, respectively. To determine the lower bound on the variance of estimator $\hat{\mathbf{\Theta}}$, one computes the Fisher information matrix $I_{F}$, given by:
\begin{flalign}
I_{F}\left(\mathbf{\Theta}\right)\,&=\,-E\left[\frac{\partial^{2}\ln{\mathcal{P}(\mathbf{X}_{s};\mathbf{\Theta})}}{\partial\mathbf{\Theta}^{2}}\right],\label{eq.1}
\end{flalign}
where $\mathcal{P}(\mathbf{X}_{s};\mathbf{\Theta})=\mathcal{P}_{s}$ is the conditional probability distribution function and $E$ is the expectation operator. The element $i,j$ of $I_{F}\left(\mathbf{\Theta}\right)$ is given by:
\begin{flalign}
I_{F}\left(\mathbf{\Theta}\right)_{ij}\,=\,-E\left[\frac{\partial^{2}\ln{\mathcal{P}(\mathbf{X}_{s};\mathbf{\Theta})}}{\partial\mathbf{\Theta}^{2}}\right]_{ij}\,=\,-\sum_{s}\frac{\partial^{2}\ln{\mathcal{P}(\mathbf{X}_{s};\mathbf{\Theta})}}{\partial\theta_{i}\partial\theta_{j}}\mathcal{P}(\mathbf{X}_{s};\mathbf{\Theta}).\label{eq.2}
\end{flalign}
The CRLB is then given by the diagonal elements of the inverse of matrix $I_{F}$, i.e.
\begin{flalign}
\text{Var}\left(\hat{\theta}_{i}\right)\,&\geq\,\left[I_{F}^{-1}\left(\mathbf{\Theta}\right)\right]_{ii},\label{eq.3}
\end{flalign}
where $\text{Var}\left(\hat{\theta}_{i}\right)$ stands for the variance of estimator $\hat{\theta}_{i}$ for the unknown parameter $\theta_{i}$. 

It is important to note that the estimator based on the maximum likelihood principle $\hat{\theta}_{ML}$ asymptotically becomes unbiased and achieves the CRLB for large data sets\cite{Kay2009}, that is:
\begin{flalign}
\hat{\mathbf{\Theta}}_{ML}\,\stackrel{a}{\sim}\,\mathcal{N}\left\lbrace \mathbf{\Theta},\textbf{diag}\left[I_{F}^{-1}\left(\mathbf{\Theta}\right)\right]\right\rbrace,\label{eq.MLE}
\end{flalign}
where $\mathcal{N}$ stands for the normal distribution and $\textbf{diag}$ takes the diagonal elements of a matrix.

\subsection{The Fisher matrix with Poisson noise in ptychography}
To find the Fisher information matrix, we start computing the second order derivative of the likelihood functional $\mathcal{L}_{P}$ with respect to $O(\textbf{r})$:
\begin{flalign}
\delta^{2}{\mathcal{L}_{P}}\left(\delta O \right)\left(\delta \tilde{O}\right)\,&=\,\frac{1}{(\hbar\omega)^{2}}\sum_{{m}}\sum_{\bm{\upxi}}\frac{n_{m}(\bm{\upxi})}{N_{m}^{2}(\bm{\upxi})}\left[\delta I_{m}\left(\delta O\right)\right]\left[\delta I_{m}\left(\delta \tilde{O}\right)\right]\nonumber\\
&\quad-\frac{1}{\hbar\omega}\sum_{{m}}\sum_{\bm{\upxi}}\left(\frac{n_{m}(\bm{\upxi})}{N_{m}(\bm{\upxi})}-1\right)\delta^{2}{I_{m}}\left(\delta O\right)\left(\delta \tilde{O}\right),\label{eq.15}
\end{flalign}
where $\delta \tilde{O}$ is the local perturbation of the value of $O$ on a discretized grid as well:
\begin{flalign}
\delta \tilde{O}\,=\,\sum_{\textbf{r}_{j}}\delta \tilde{O}(\textbf{r}_{j})\delta(\textbf{r}-\textbf{r}_{j})\,=\,\sum_{\textbf{r}_{j}}
\begin{bmatrix}
\delta \tilde{A}(\textbf{r}_{j})\\
\text{i}\tilde{A}(\textbf{r}_{j})\delta \tilde{\phi}(\textbf{r}_{j})
\end{bmatrix}
e^{\text{i}\tilde{\phi}(\textbf{r}_{j})}\delta(\textbf{r}-\textbf{r}_{j})\label{eq.27}.
\end{flalign}
By taking the expectation of Eq. (\ref{eq.15}), we get:
\begin{flalign}
E\left(\delta^{2}{\mathcal{L}_{P}}\right)\left(\delta O \right)\left(\delta \tilde{O}\right)\,&=\,\frac{1}{(\hbar\omega)^{2}}\sum_{{m}}\sum_{\bm{\upxi}} E\left\lbrace\frac{n_{m}(\bm{\upxi})}{N_{m}^{2}(\bm{\upxi})}\left[\delta I_{m}\left(\delta O\right)\right]\left[\delta I_{m}\left(\delta \tilde{O}\right)\right]\right\rbrace\nonumber\\
&\quad-\frac{1}{\hbar\omega}\sum_{{m}}\sum_{\bm{\upxi}} E\left[\left(\frac{n_{m}(\bm{\upxi})}{N_{m}(\bm{\upxi})}-1\right)\delta^{2}{I_{m}}\left(\delta O\right)\left(\delta \tilde{O}\right)\right],\label{eq.16}
\end{flalign}
in which we commute the expectation and summation because the measurements $n_{m}(\bm{\upxi})$ are independent photon measurements. Using the properties of the Poisson distribution\cite{Fienup1993}:
\begin{align}
\left\lbrace 
\begin{tabular}{l}
$\sum\limits_{n_{m}}\dfrac{N_{m}^{n_{m}}}{n_{m}!}e^{-N_{m}}\,=\,1$,\\
$\sum\limits_{n_{m}}n_{m}\dfrac{N_{m}^{n_{m}}}{n_{m}!}e^{-N_{m}}\,=\,N_{m}$,
\end{tabular}
\right.\label{eq.17}
\end{align}
and using Eq. (\ref{eq.11}), we find:
\begin{flalign}
E\left(\delta^{2}{\mathcal{L}_{P}}\right)\left(\delta O \right)\left(\delta \tilde{O}\right)\,&=\,\frac{1}{(\hbar\omega)^{2}}\sum_{{m}}\sum_{\bm{\upxi}}\frac{1}{N_{m}(\bm{\upxi})}\delta I_{m}\left(\delta O\right)\delta I_{m}\left(\delta \tilde{O}\right)\nonumber\\
&=\,\frac{4}{\hbar\omega}\sum_{{m}}\sum_{\bm{\upxi}}\frac{1}{I_{m}(\bm{\upxi})}\Re\left[\mathcal{F}\left( \varPsi_{m}\right)\mathcal{F}\left( P_{m}\delta O\right)^{*}\right]\Re\left[\mathcal{F}\left(\varPsi_{m} \right)\mathcal{F}\left( P_{m}\delta \tilde{O}\right)^{*}\right] \nonumber\\
&=\,\frac{2}{\hbar\omega}\sum_{{m}}\sum_{\bm{\upxi}}\Re\left[\frac{\left[\mathcal{F}\left( \varPsi_{m}\right)(\bm{\upxi})\right]^{2}}{I_{m}(\bm{\upxi})}\mathcal{F}\left( P_{m}\delta O\right)^{*}\mathcal{F}\left( P_{m}\delta \tilde{O}\right)^{*}\right]\nonumber\\
&\quad+\frac{2}{\hbar\omega}\sum_{{m}}\sum_{\bm{\upxi}}\Re\left[\mathcal{F}\left( P_{m}\delta O\right)\mathcal{F}\left( P_{m}\delta \tilde{O}\right)^{*}\right].\label{eq.18}
\end{flalign}

From Eq. (\ref{eq.23}), Eq. (\ref{eq.27}) and Eq. (\ref{eq.18}) we can derive the discretized Fisher information matrix with respect to the transmission and the thickness function of the object:
\begin{flalign}
I_{F,ij}\,=\,
\begin{bmatrix}
(I_{F})_{AA,ij}&(I_{F})_{A\phi,ij}\\
(I_{F})_{\phi A,ij}&(I_{F})_{\phi \phi,ij}
\end{bmatrix}\nonumber
&=\,
\frac{2}{\hbar\omega}\sum_{{m}}
\begin{bmatrix}
\Re\left[f_{m}(\textbf{r}_{i},\textbf{r}_{j})\right]&\Im\left[A(\textbf{r}_{j})f_{m}(\textbf{r}_{i},\textbf{r}_{j})\right]\\
\Im\left[A(\textbf{r}_{i}) f_{m}(\textbf{r}_{i},\textbf{r}_{j})\right]&-\Re\left[A(\textbf{r}_{i})A(\textbf{r}_{j})f_{m}(\textbf{r}_{i},\textbf{r}_{j})\right]
\end{bmatrix}\nonumber\\
&\quad\,+
\frac{2}{\hbar\omega}\sum_{{m}}
\begin{bmatrix}
\left|P_{m}(\textbf{r}_{i})\right|^{2}\delta_{ij}&0\\
0&A^{2}(\textbf{r}_{i})\left|P_{m}(\textbf{r}_{i})\right|^{2}\delta_{ij}
\end{bmatrix}
,\label{eq.21}
\end{flalign}
where the auxiliary function$f$ is given by:
\begin{flalign}
f_{m}(\textbf{r}_{i},\textbf{r}_{j})\,&=\,\mathcal{F}^{-1}\left[\frac{\mathcal{F}\left( \varPsi_{m}\right)}{\mathcal{F}\left( \varPsi_{m}\right)^{*}}\right](\textbf{r}_{i}+\textbf{r}_{j})\cdot P_{m}^{*}(\textbf{r}_{i})P_{m}^{*}(\textbf{r}_{j})e^{-\text{i}\left[\phi(\textbf{r}_{i})+\phi(\textbf{r}_{j})\right]}.\label{eq.22}
\end{flalign}
where we used Eq. (\ref{eq.6}) and Kronecker's symbol $\delta_{ij}$.

In Eq. (\ref{eq.21}) we see that the first term is symmetric and the second one is diagonal. The analytical expression for the CRLB, which is obtained from the inverse of the Fisher matrix, cannot be easily derived, but this inverse can be computed numerically. Detailed examples are presented in the next section.

\section{Direct calculation of the CRLB}

As shown in Eq. (\ref{eq.3}), the CRLB is given by the diagonal elements of the inverse of matrix $I_{F}$, which can be obtained by numerical computations. In this section, we present the results of some computed CRLB. To investigate how the illumination (i.e. the probe function $P$) and the object $O$ influence the CRLB, we study four cases separately, as described in Table \ref{table.1}. Note that only Poisson noise is applied throughout our simulations. Other noise models (e.g. Gaussian noise or Poisson-Gaussian noise \cite{Zhang2017}) should be included when these are dominant. All of the calculation results given in this section are compared to the Monte Carlo experiment result that are presented in the next section.
\begin{table}
	\centering
	\caption{Four cases that are considered in the computation of the CRLB}
	\begin{tabular}{c|l}
		\hline
		{\color{cyan}{\textbf{Case-1}}} & \makecell[l]{Both the transmission and thickness function of the object are uniform.
		The probe \\ has structured wavefront but uniform illumination power in the circular support.} \\
		\hline
		{\color{cyan}{\textbf{Case-2}}} & \makecell[l]{Both the transmission and thickness function of the object are uniform.
		The probe \\ has structured wavefront and structured illumination power in the circular support.} \\
		\hline
		{\color{cyan}{\textbf{Case-3}}} & \makecell[l]{The object has non-uniform transmission but uniform thickness function. \\
		The probe is a plane-wave with circular support.}\\
		\hline
		{\color{cyan}{\textbf{Case-4}}} & \makecell[l]{The object has uniform transmission but non-uniform thickness function. \\
		The probe is a plane-wave with circular support.} \\
		\hline
	\end{tabular}
	\label{table.1}
\end{table}

For all cases shown in Table \ref{table.1}, the probe moves over the object by a $2\times2$ regular grid. In line with the conventional ptychography configuration, the overlap ratio between adjacent illuminated areas is 70\%, which is regarded as \textit{a prior} knowledge and employed in the reconstruction algorithm. The overlap ratio is defined as follows. Suppose the diameter of the circular support is $L$, and the distance between corresponding points in adjacent illumination positions is $d$, where $0<d<L$. The overlap ratio is then defined by:
\begin{align}
\text{overlap ratio}\,=\,1-\frac{d}{L}\label{overlap}
\end{align}
which is usually chosen between 60\% and 85\% to achieve optimal performance of the reconstruction algorithm\cite{Bunk2008}.
 
The characteristic parameters for the numerical computations are shown in Table \ref{table.2}. The object is discretised and zero padded by a $70\times70$ square grid with grid spacing $1\mu m$. The total illuminated area is roughly $40\times40\mu m^{2}$. The circular probe has radius $30\mu m$ and is discretised by a square grid of $60\times60$ grid points with grid spacing of $1\mu m$. The wavelength is 30 nm. The far field intensities are measured with a detector at propagation distance of $5cm$ behind the object. The detector consists of an array of $60\times60$ pixels with pixel size $50\mu m$. Hence the maximum spatial frequency (without factor $2\pi$) that is measured is $1\mu m^{-1}$ and the frequency are sampled with distance $\frac{1}{30} \mu m^{-1}$.  
\begin{table}
	\centering
	\caption{The characteristic parameters for the simulations}
	\begin{tabular}{||c|c|c|p{0.1cm}|p{1.1cm}|c|c|c||}
		\hhline{|t:========:t|}
		\multirow{2}{*}{probe} & grid size & \makecell[c]{grid\\spacing} & \multicolumn{2}{c|}{wavelength} & \makecell[c]{scanning\\grid} & \makecell[c]{overlap\\ratio} & \makecell[c]{radius of\\circular support} \\
        \hhline{|~-------||}
        & $60\times60$ & $1\mu m$ & \multicolumn{2}{c|}{$30nm$} & $2\times2$ & $70\%$ & $30\mu m$\\
        \cline{4-4}\hhline{|:===~====:|}
		\multirow{2}{*}{object} & grid size & \makecell[c]{grid\\spacing} & & \multirow{2}{*}{detector} & \makecell[c]{pixel\\number}& pixel size & \makecell[c]{propagation\\distance} \\
        \hhline{|~--~~---||}
		& $70\times70$ & $1\mu m$ & & & $60\times60$ & $50\mu m$ & $5cm$ \\
		\hhline{|b:===b-b====:b|}
	\end{tabular}
	\label{table.2}
\end{table}

To compute the CRLB, we first construct the Fisher information matrix $I_{F}$ using Eq. (\ref{eq.21}). Although the number of degrees of freedom used to describe the object is small, namely $70\times70\times2$ elements, where the factor 2 is due to the fact that the object function is complex, the discretised Fisher matrix already includes $9800\times9800$ elements. The CRLB is obtained by numerically computing the inverse of $I_{F}$. Since $I_{F,ij}$ is an symmetric matrix with real entries, one can apply the eigenvalue decomposition to find the inverse of the Fisher matrix. We select the eigenvalues of $I_{F}$ that are bigger than a default tolerance, then use these eigenvalues and the corresponding eigenvectors to compute the inverse of $I_{F}$. This calculation is done by utilizing the 'pinv' routine in MATLAB. The diagonal elements of the inverse matrix $I^{-1}_{F}$ consists of an array of $70\times70\times2$ elements, of which the first $70\times70$ elements correspond to the CRLB of $A(\textbf{r})$ and the last $70\times70$ elements contain the CRLB of $\phi(\textbf{r})$ .

We define the illumination power by means of the total photon number (PN) counting over the cross section of the probe, given by: 
\begin{align}
\text{PN}\,=\,\dfrac{\sum_{\textbf{r}}\left|P(\textbf{r})\right|^{2}}{\hbar\omega}.\label{PN}
\end{align}
An important property of the CRLB is that it is proportional to the reciprocal of the illumination power. This property follows from the fact that Eq. (\ref{eq.21}) and Eq. (\ref{eq.22}) are proportional to the square of the input power. The observation that the CRLB scales with the reciprocal of the illuminating power is confirmed by the computations discussed below. 

In the remainder of this section we show the computed CRLB for high illumination power, i.e. $\text{PN}=10^{9}$, and for low illumination power, i.e. $\text{PN}=10^{3}$, as examples. The influence of the object and the probe on the CRLB will be discussed separately.

\subsection{The influence of the illumination on the CRLB}
\begin{figure}[htp!]
	\centering\includegraphics[width=0.95\textwidth]{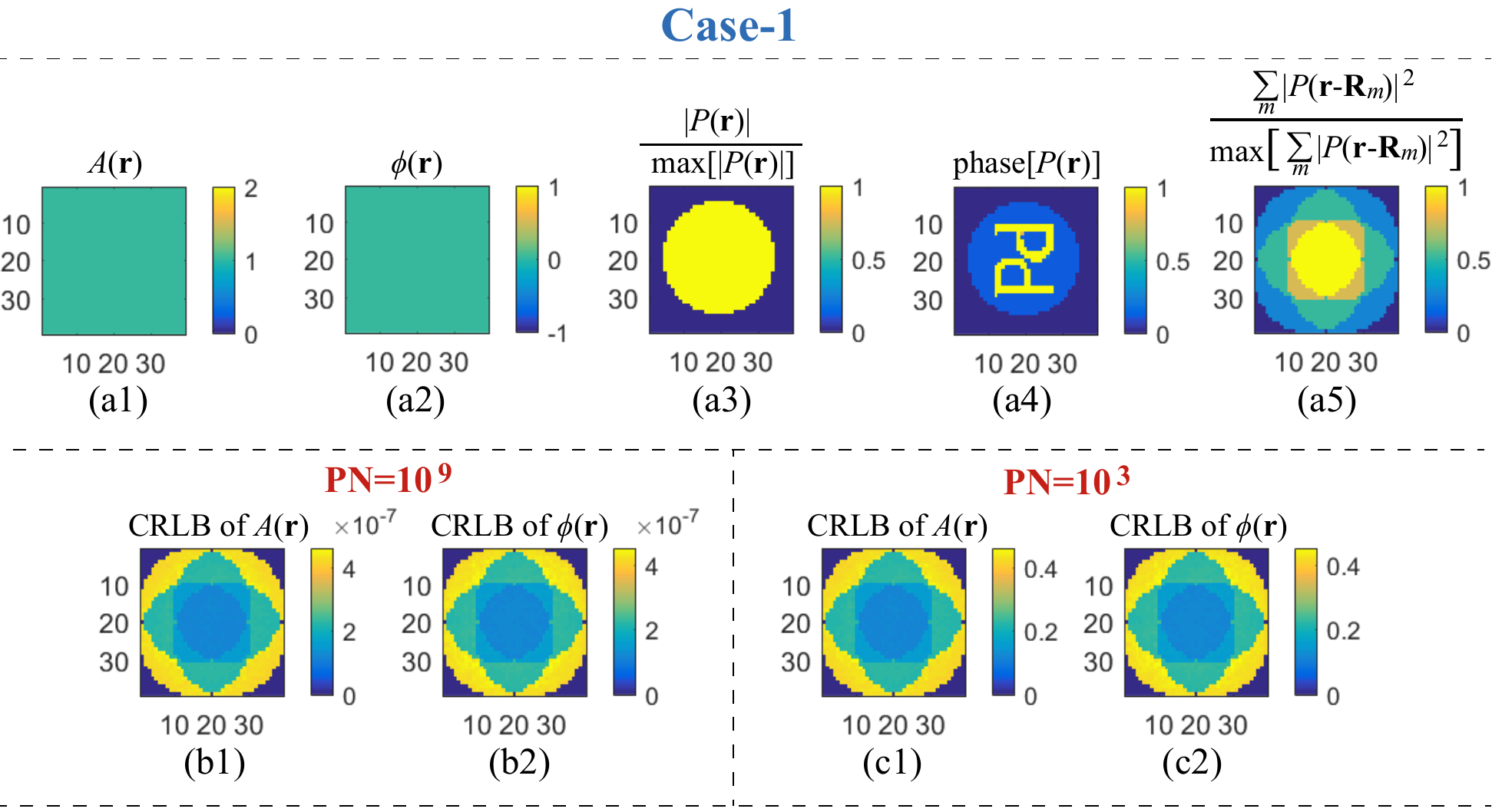}	\caption{The CRLB computed from the Fisher matrix for Case-1. (a1) and (a2) are the object's actual transmission $A(\textbf{r})$ and actual phase function $\phi(\textbf{r})$, respectively. (a3) and (a4) show the actual amplitude and phase of the probe function, respectively. (a5) shows the normalized sum of the intensities of the illuminations. (b1) and (b2) show the CRLB of $A(\textbf{r})$ and $\phi(\textbf{r})$, respectively, for the case of $\text{PN}=10^{9}$. (c1) and (c2) are the CRLB for the case of $\text{PN}=10^{3}$.}\label{Fig.1}
\end{figure}

\begin{figure}[htp!]
	\centering\includegraphics[width=0.95\textwidth]{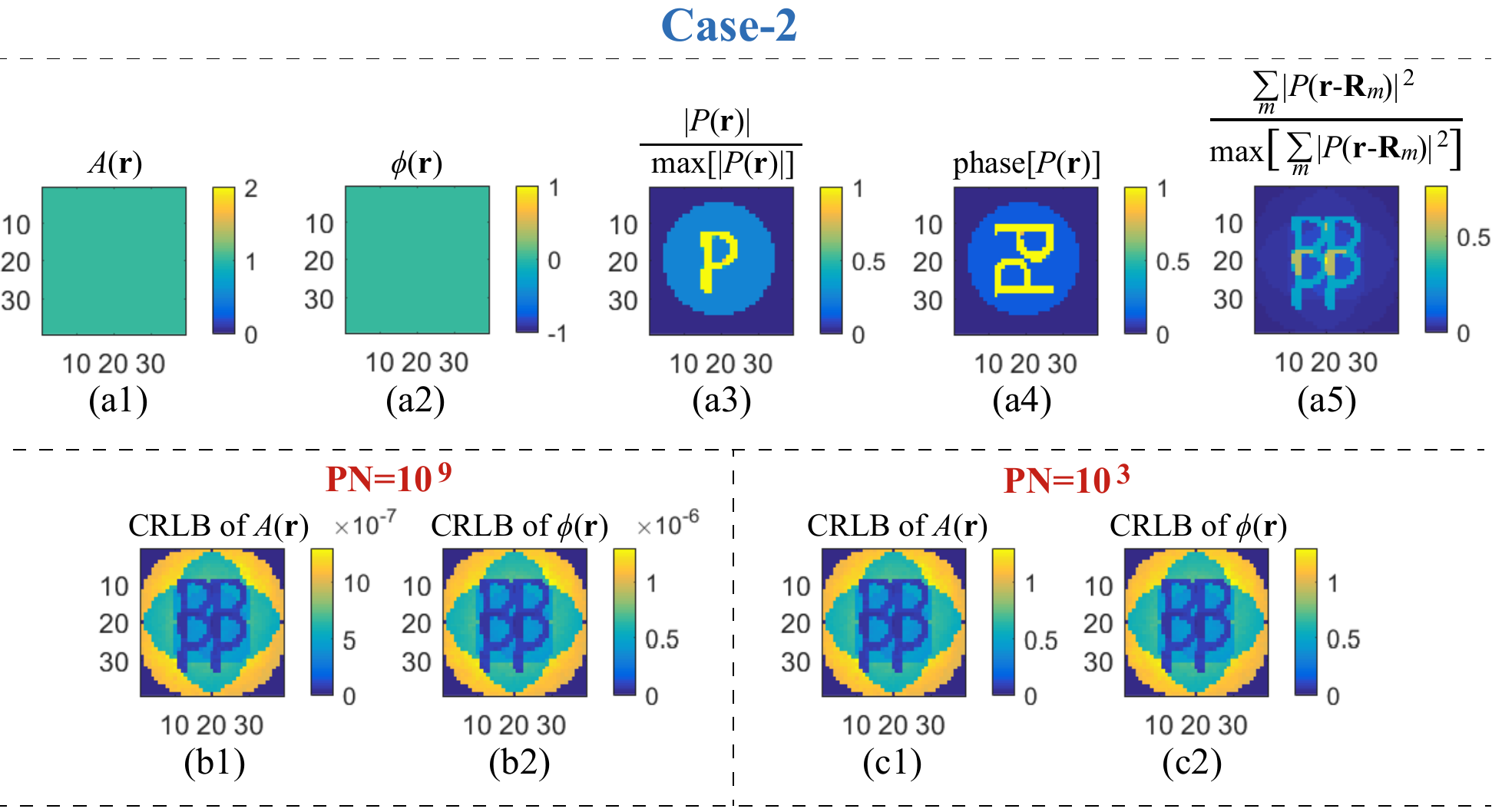}	\caption{The calculated CRLB for Case-2. (a1) - (a5) are the actual object, probe and the normalized sum of the intensities of the illuminations, respectively. (b1) and (b2) are the CRLB of $A(\textbf{r})$ and $\phi(\textbf{r})$, respectively, for $\text{PN}=10^{9}$. (c1) and (c2) are the CRLB for the case of $\text{PN}=10^{3}$.}\label{Fig.2}
\end{figure}

In order to investigate the influence of the illumination on the CRLB, we start by studying Case-1 and Case-2 described in Table \ref{table.1}. For these cases, the actual object, the actual illumination and the computed CRLB are shown in  Fig. \ref{Fig.1} and Fig. \ref{Fig.2}.
We let the object have uniform transmission and thickness function for the time being. For Case-1, the probe function $P$ has uniform power throughout its circular support and zero value outside its support, but the phase of the probe has variation in the form of two characters 'P' as shown in Fig. 1a4. On the other hand, the illumination in Case-2 has the shape of the character 'P' and truncated by the circular support as shown in Fig. \ref{Fig.2}a3, and its phase has the same features consisting of two characters 'P' as in Case-1 (see Fig. 2a3 and Fig. 2a4). Considering that a perfectly collimated beam is difficult to obtain, we have chosen the wavefront of the illumination to be non-uniform for both Case-1 and Case-2.

It is seen in Fig. \ref{Fig.1} that the CRLB of the object resembles normalized sum of the intensities of the illuminations shown in Fig. \ref{Fig.1}a5. In particular, the part of the object which is illuminated 4 times reaches a variance approximately 4 times smaller than the part which is illuminated only once, and this conclusion holds for both the object's local transmission $A(\textbf{r})$ and phase function $\phi(\textbf{r})$. Interestingly, when the dose distribution of the illumination is more complicated as given in Fig. \ref{Fig.2}a3 and Fig. \ref{Fig.2}a4, the CRLB shown in Fig. \ref{Fig.2}b and Fig. \ref{Fig.2}c again resemble the overall illumination pattern shown in Fig. \ref{Fig.2}a5. In other words, the more illumination power we apply to the object, the lower the minimum variance of the obtained reconstruction. One can notice that the maximum of the CRLB in Fig. \ref{Fig.2}c and Fig. \ref{Fig.2}d is in the yellow corner and is larger than the CRLB in Fig. \ref{Fig.1}. This is because for Case-2 the illuminating power is concentrated in the 'P' character, as shown in Fig. \ref{Fig.2}(a3). Around the yellow corner there are parts of the object where the computed CRLB is zero. These parts of the object are not illuminated. For the areas where $I_{F}$ is zero, the computed CRLB is also equal to zero because we ignore the singular values of $I_{F}$. In reality the CRLB there is  infinite.

Moreover, we can see in Fig. \ref{Fig.1} and Fig. \ref{Fig.2} that the CRLB is linearly proportional to the inverse of $\text{PN}$ (i.e. the illumination power). This calculation result is in agreement with Eq. (\ref{eq.21}) because the probe function $P(\textbf{r})$ can be written as the factor $\sqrt{\text{PN}}$ times the normalized $P(\textbf{r})$. On the other hand, the computed CRLB of both $A(\textbf{r})$ and $\phi(\textbf{r})$ do not show any influence due to the spatial variation of the phase of the probe. Therefore, we conclude that it is the illumination intensity pattern, i.e. the dose distribution, which strongly determines the CRLB in ptychography for Poisson noise.

\subsection{The influence of the object on the CRLB}
\begin{figure}[htp!]
	\centering\includegraphics[width=0.95\textwidth]{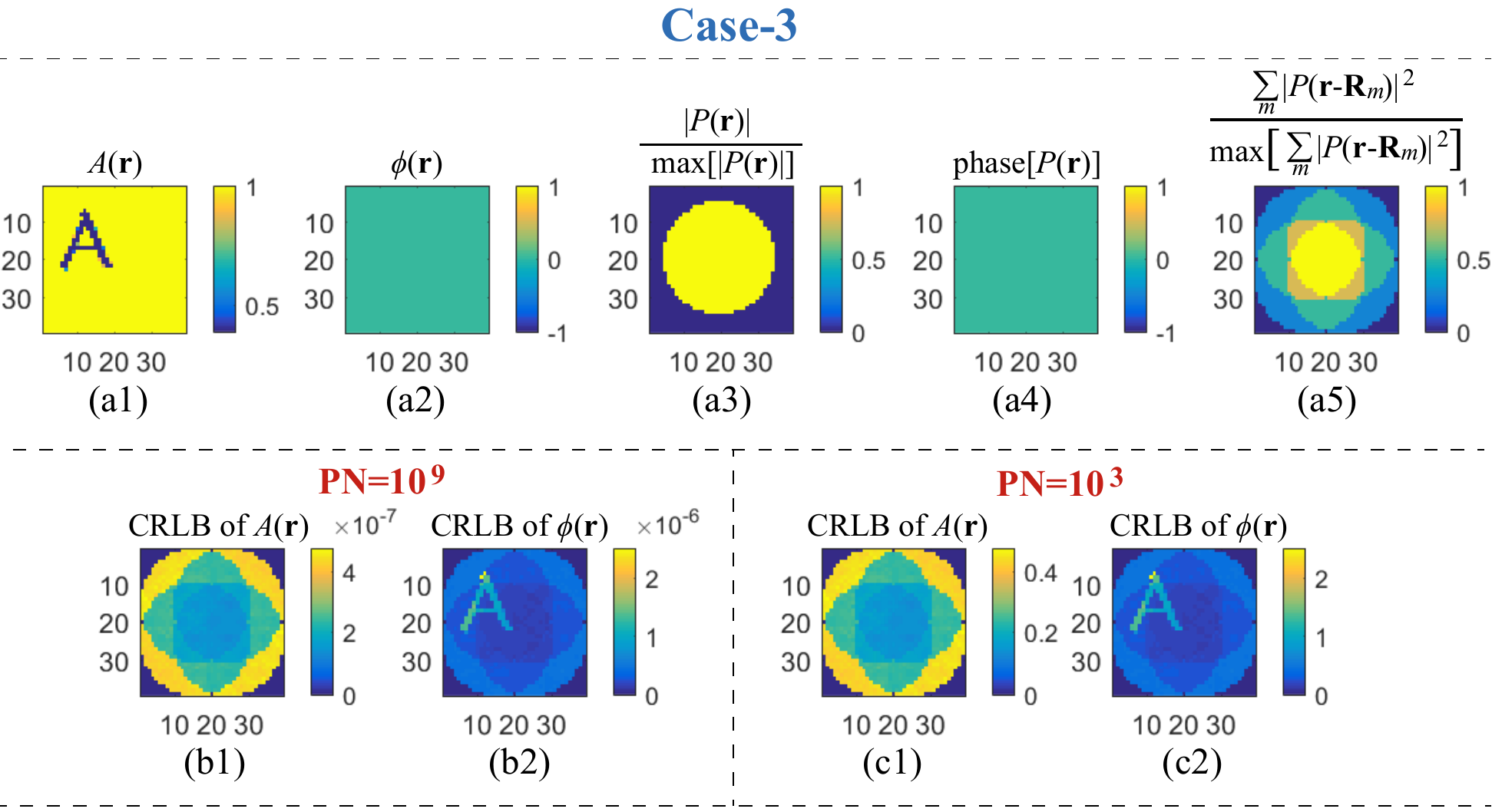}	\caption{The CRLB for Case-3. (a1) - (a5) are the actual object, probe and the normalized sum of the intensities of the illuminations, respectively. (b1) and (b2) are the CRLB of $A(\textbf{r})$ and $\phi(\textbf{r})$, respectively, for $\text{PN}=10^{9}$. (c1) and (c2) are the CRLB when $\text{PN}=10^{3}$.}\label{Fig.3}
\end{figure}

\begin{figure}[htp!]
	\centering\includegraphics[width=0.95\textwidth]{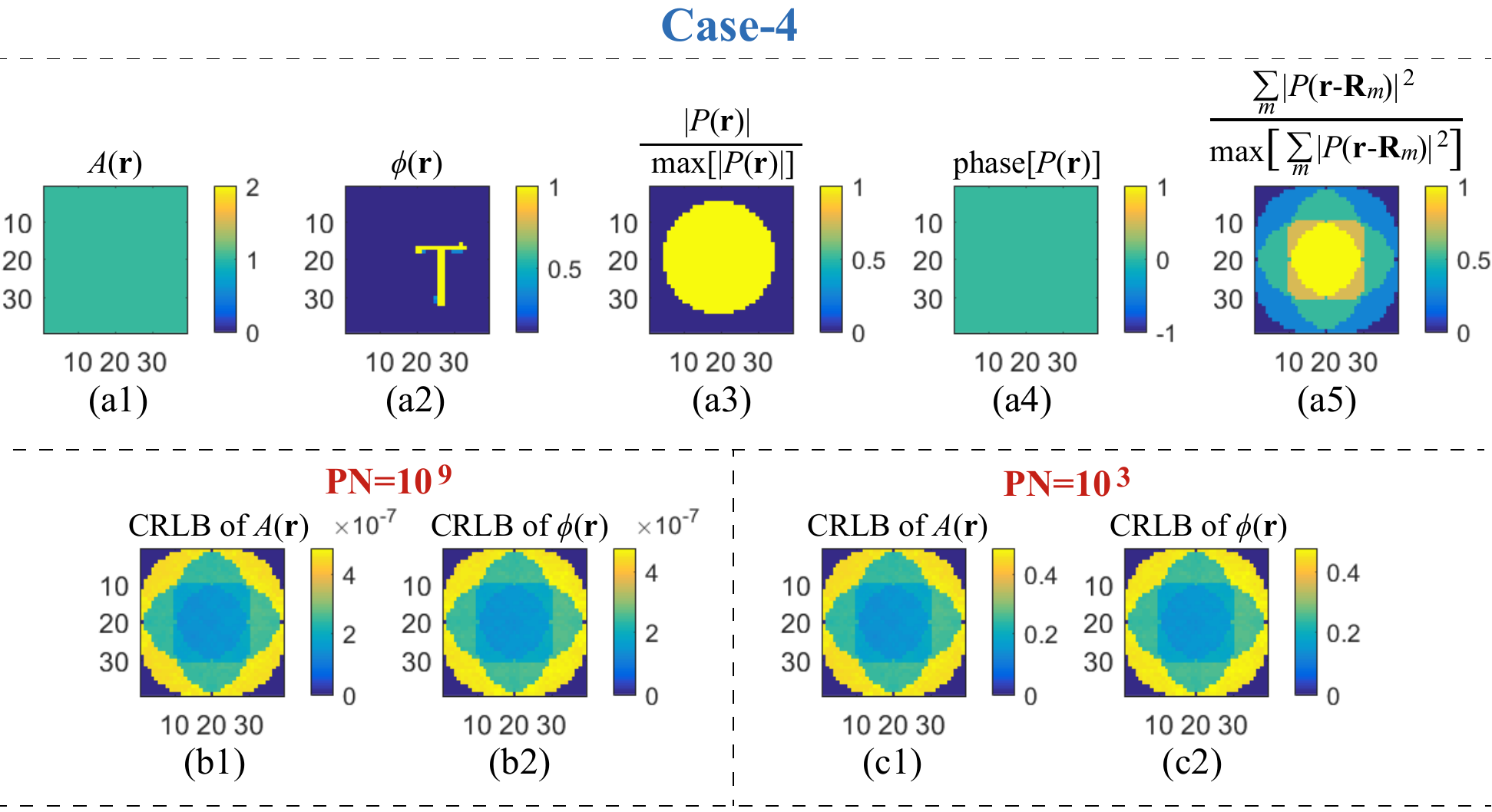}	\caption{The CRLB for Case-4. (a1) - (a5) are the actual object, probe and the normalized sum of the intensities of the illuminations, respectively. (b1) and (b2) are the CRLB of $A(\textbf{r})$ and $\phi(\textbf{r})$, respectively, when  $\text{PN}=10^{9}$. (c1) and (c2) are the CRLB for $\text{PN}=10^{3}$.}\label{Fig.4}
\end{figure}

The Fisher matrix in Eq. (\ref{eq.21}) is in fact a function of the object, and hence so is the CRLB. To find the influence of $A(\textbf{r})$ and $\phi(\textbf{r})$ on the CRLB, we focus on Case-3 and Case-4 from now on. To reduce the influence of the illumination to a minimum, we let the probe function be a plane-wave with circular support. The influence of the object's transmission and phase function is investigated separately. In Case-3 we let the function $A(\textbf{r})$ have the shape of the character 'A' while $\phi(\textbf{r})$ is kept uniform, as shown in Fig. \ref{Fig.3}. The minimum value of $A(\textbf{r})$ is 0.1. For Case-4, the function $A(\textbf{r})$ is uniform whereas the phase function $\phi(\textbf{r})$ has the shape of the character 'T' as shown in Fig. \ref{Fig.4}.

The computed CRLB of the object for Case-3 and Case-4 is illustrated in Fig. \ref{Fig.3}b, Fig. \ref{Fig.3}c, Fig. \ref{Fig.4}b and Fig. \ref{Fig.4}c, respectively. It is clear that our conclusion in Section 3.1 still holds, i.e. the CRLB is very similar to the pattern of the sum of the intensities of the illuminations. On the other hands, we can see also that the object's local transmission $A$ is predominate in determining the CRLB of $\phi$, as shown in Fig. \ref{Fig.3}(b2) and Fig. \ref{Fig.3}(c2). This result agrees with Eq. (\ref{eq.21}), because the function $A$ appears in the terms of $I_{F}$ which relates to $\phi$. However, the influence of $\phi$ on the CRLB is much less than $A$. Therefore, we conclude that the second term in Eq. (\ref{eq.21}) dominant. In other words, when the estimator of ptychography is unbiased, the variance of the object's transmission $A(\textbf{r})$ is strongly determined by the illumination power and dose distribution, whereas the variance of the object's phase $\phi(\textbf{r})$ is influenced by both of the transmission $A(\textbf{r})$, the illumination power and the dose distribution.

In the next section, the CRLB shown in Fig. \ref{Fig.1} - Fig. \ref{Fig.4} are used as references for Monte Carlo experiments.

\section{Monte Carlo analysis}
To validate our calculation of the CRLB, Monte Carlo computations have been performed. For consistency, we discretise the probe and the object in the same way as described in Table \ref{table.2}. The wavelength, object, probe, far field measurements and grid sizes are as described in Table \ref{table.2} also. The Fresnel number of the system is 0.15. Hence for this configuration the detector is in the Fraunhofer region. 

The ptychographic data with various level of noise is generated as follows. For every ptychography simulation and for every probe position, we first assign the probe function with corresponding photon numbers in accordance with the $\text{PN}$ that is chosen. Then, the noise-free diffracted wavefield in the far field is calculated, and the Poisson random number generator in MATLAB is applied to generate the noisy data. 

To verify the asymptotic property of the maximum likelihood method of Eq. (\ref{eq.MLE}), we developed and implemented \textbf{Algorithm \ref{alg.1}} as described in the Appendix. To mitigate ambiguity problems of ptychography\cite{Wei2019}, e.g. the global phase shift, the conjugate reconstruction and the raster grid pathology, it is assumed that the probe used in the Monte Carlo experiment is known. To shorten the computation time and to improve the convergence, the conjugate gradient method \cite{Thibault2012,Tripathi2014} is implemented in \textbf{Algorithm \ref{alg.1}}.

For comparison, the performance of another popular method, namely the amplitude-based cost function minimization approach \cite{Guizar-Sicairos2008}, was investigated in the Monte Carlo experiment also. This is implemented in \textbf{Algorithm \ref{alg.2}}. The idea of this algorithm is to retrieve the object by minimizing the cost function defined in Eq. (\ref{eq.8}). We remark that one can alternatively derive \textbf{Algorithm \ref{alg.2}} from the maximum likelihood method by using the variance stabilization transform \cite{Bartlett1936,Anscombe1948,Godard2012,Thibault2012,Zhang2017}. \textbf{Algorithm \ref{alg.2}} is also described in the Appendix.

To investigate the performance of the above mentioned algorithms, the variance and the squared bias of the estimator are evaluated in our Monte Carlo analysis. Explicitly, the variance of an estimator $\hat{O}(\textbf{r})$ is defined by \cite{Kay2009}:
\begin{align}
\text{Var}\left[\hat{O}(\textbf{r})\right]\,=\,E\left\lbrace\left[\hat{O}(\textbf{r})-\left\langle \hat{O}(\textbf{r})\right\rangle\right]^{2}\right\rbrace,\qquad \text{where}\quad
\left\langle \hat{O}(\textbf{r})\right\rangle\,=\,E\left[\hat{O}(\textbf{r})\right],\label{var}
\end{align}
and the squared bias of the estimator is given by:
\begin{align}
\text{Bias}^{2}\left[\hat{O}(\textbf{r})\right]\,=\,\left|\left\langle \hat{O}(\textbf{r})\right\rangle-O_{\text{o}}(\textbf{r})\right|^{2},\label{mse-var-bias}
\end{align}
where $O_{\text{o}}$ is the actual object function.

In order to compute the expectation accurately, 2000 individual ptychographic data sets have been generated for all for cases mentioned in Table \ref{table.1} and for different value of $\text{PN}$. These data-sets have been post-processed by \textbf{Algorithm \ref{alg.1}} and \textbf{Algorithm \ref{alg.2}}, respectively, and the results are discussed next.

\subsection{The statistic properties of Maximum likelihood method and Amplitude-based cost minimization method, and the influence of the illumination}
We begin with the case of uniform object function and structured illumination, i.e. Case-1 and Case-2. For these cases the actual object and probe function are as in Fig. \ref{Fig.1}a and Fig. \ref{Fig.2}a. 

\begin{figure}[htp!]
	\centering\includegraphics[width=0.9\textwidth]{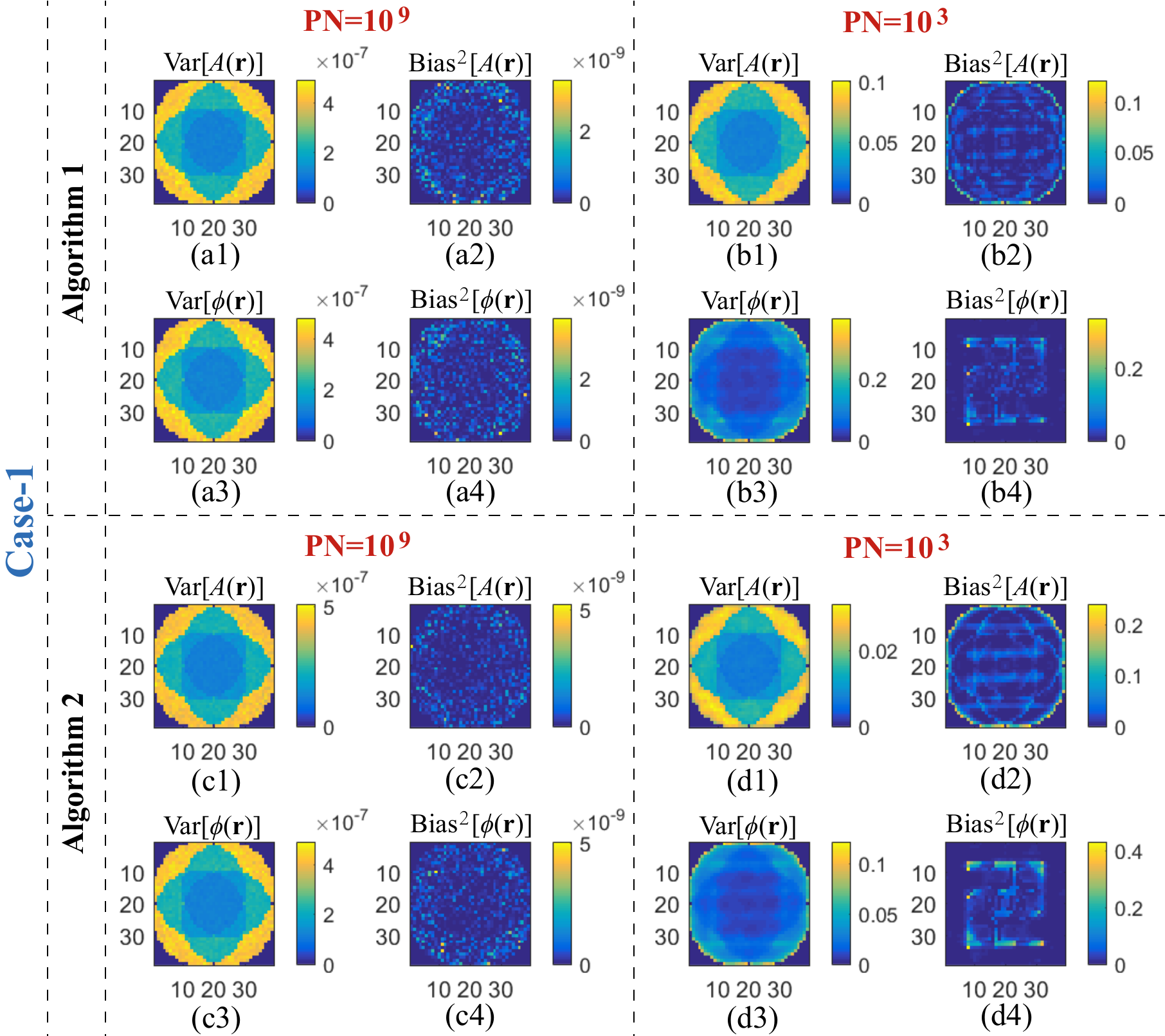}	
	\caption{The result of Monte Carlo experiment for Case-1. (a1) and (a2) are the variance and bias squared of the object's transmission $A$ when $\text{PN}=10^{9}$, respectively, obtained with \textbf{Algorithm \ref{alg.1}}. (a3) and (a4) are the variance and bias squared of the object's thickness $\phi$, respectively. (b1)-(b4) show the variance and bias squared when $\text{PN}=10^{3}$, respectively, obtained with \textbf{Algorithm \ref{alg.1}}. (c1)-(c4) and (d1)-(d4) show the results obtained with Algorithm 2 when $\text{PN}=10^{9}$ and $\text{PN}=10^{3}$, respectively.}\label{Fig.5}
\end{figure}

\begin{figure}[htp!]
	\centering\includegraphics[width=0.9\textwidth]{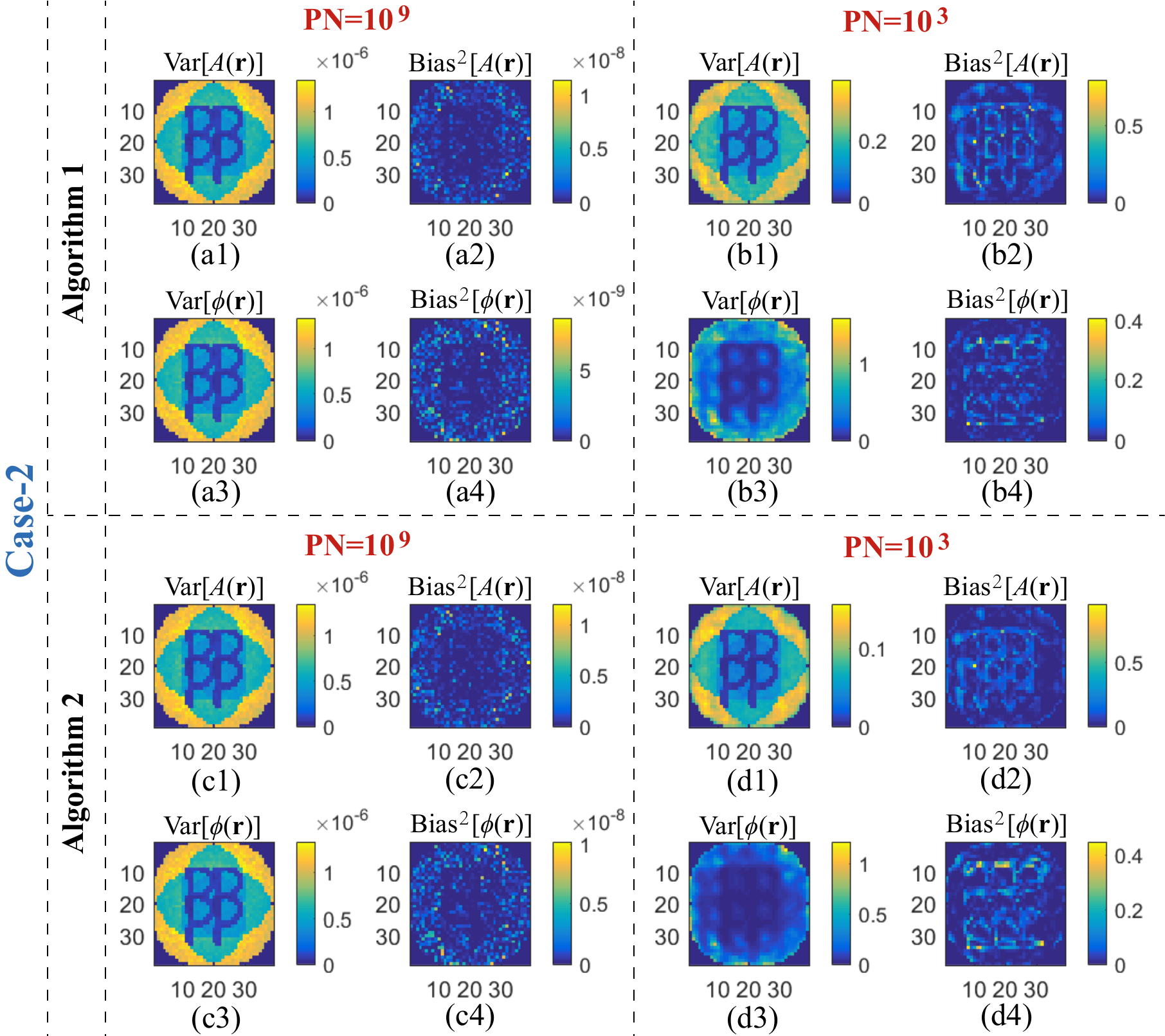}	
	\caption{The Monte Carlo experiment result for Case-2.}\label{Fig.6}
\end{figure}

When the illumination have a uniform dose distribution but a structured wavefront, the variance and bias of both \textbf{Algorithm \ref{alg.1}} and \textbf{Algorithm \ref{alg.2}} are shown in Fig. \ref{Fig.5}. 
In line with the CRLB given in \ref{Fig.1}b, we see that both algorithms that asymptotically achieve the CRLB when $\text{PN}=10^{9}$. The squared bias of the two algorithms are 100 times smaller than the variance, hence both \textbf{Algorithm \ref{alg.1}} and \textbf{Algorithm \ref{alg.2}} are asymptotically unbiased when the photon number is high. Meanwhile, by inspecting Fig. \ref{Fig.5}a and Fig. \ref{Fig.5}c, one can infer that the variance of both algorithms are related to the local illuminating power as mentioned in Section 3.2, i.e. the parts of the object that are illuminated 4 times have a variance that is 4 times smaller than the parts that are illuminated only once. A very similar conclusion can be made for Case-2, i.e. when the illumination's local dose distribution is not uniform. As shown in Fig. \ref{Fig.6}a and Fig. \ref{Fig.6}c, the variance of both algorithms agree with the CRLB given in Fig. \ref{Fig.2}b and is inversely proportional to the local illumination power given in Fig. \ref{Fig.2}a5.

When the photon number is low, i.e. $\text{PN}=10^3$, \textbf{Algorithm \ref{alg.1}} and \textbf{Algorithm \ref{alg.2}} behave differently with the current data-set. In particular, we see in Fig. \ref{Fig.5} and Fig. \ref{Fig.6} that \textbf{Algorithm \ref{alg.1}} in fact reaches smaller bias than \textbf{Algorithm \ref{alg.2}} when the photon number is low. This suggests that the approach based on the maximum likelihood principle can provide less bias than the amplitude-based cost function minimization method. Meanwhile, the variance of the estimator \textbf{Algorithm \ref{alg.2}} tends to be smaller than \textbf{Algorithm \ref{alg.1}}. This can be explained from the fact that minimizing the amplitude-based cost function minimization can approximately be regarded as a variance stabilizing de-noising algorithm \cite{Bartlett1936,Godard2012,Thibault2012,Zhang2017}. On the other hand, the two algorithms share certain properties. For low photon count, both \textbf{Algorithm \ref{alg.1}} and \textbf{Algorithm \ref{alg.2}} have lower variance than the CRLB, which indicates they cannot converge to unbiased estimators and cannot reach the CRLB with the current Monte Carlo data-set. More discussion about this slow convergence is given in Section 4.4.

In Fig. \ref{Fig.5} and Fig. \ref{Fig.6} we see that the wavefront profile of the probe only appears in the bias of the reconstruction when the photon count is low. The local illumination power determines the bias for Case-3 and Case-4 for $\text{PN}=10^{3}$ as well. For higher photon number, e.g. $\text{PN}=10^9$, there is no trace of the illumination in the bias for Case-3 and only negligible trace of illumination's local power for Case-4. Therefore, we conclude that the illumination's wavefront profile only influence the statistic property of the algorithms when the photon count is low, whereas the illumination's local power always influences the variance.

\subsection{The influence of the object on the variance and bias}
\begin{figure}[htp!]
	\centering\includegraphics[width=0.9\textwidth]{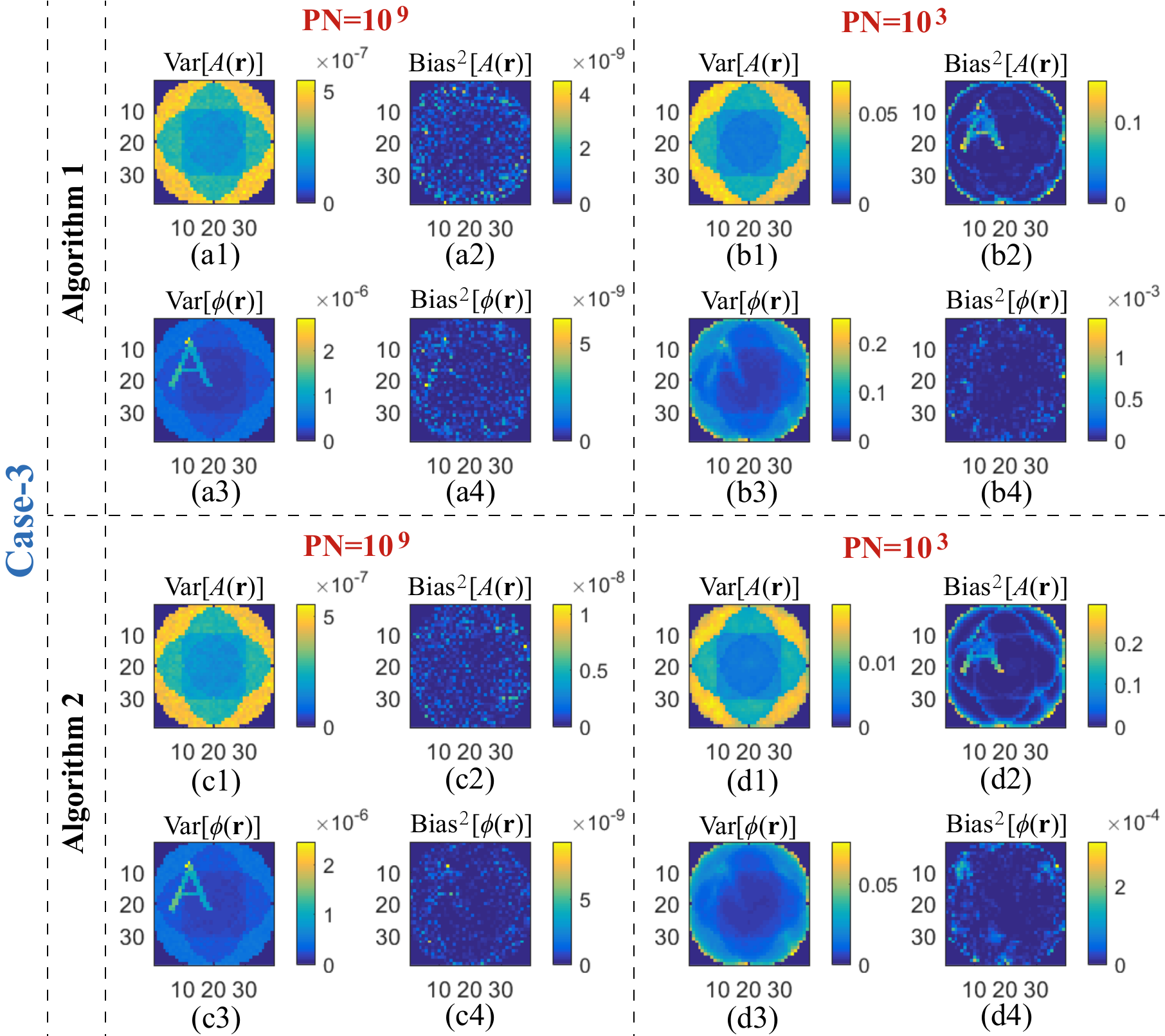}
	\caption{The Monte Carlo experiment result for Case-3.}\label{Fig.7}
\end{figure}

\begin{figure}[htp!]
	\centering\includegraphics[width=0.9\textwidth]{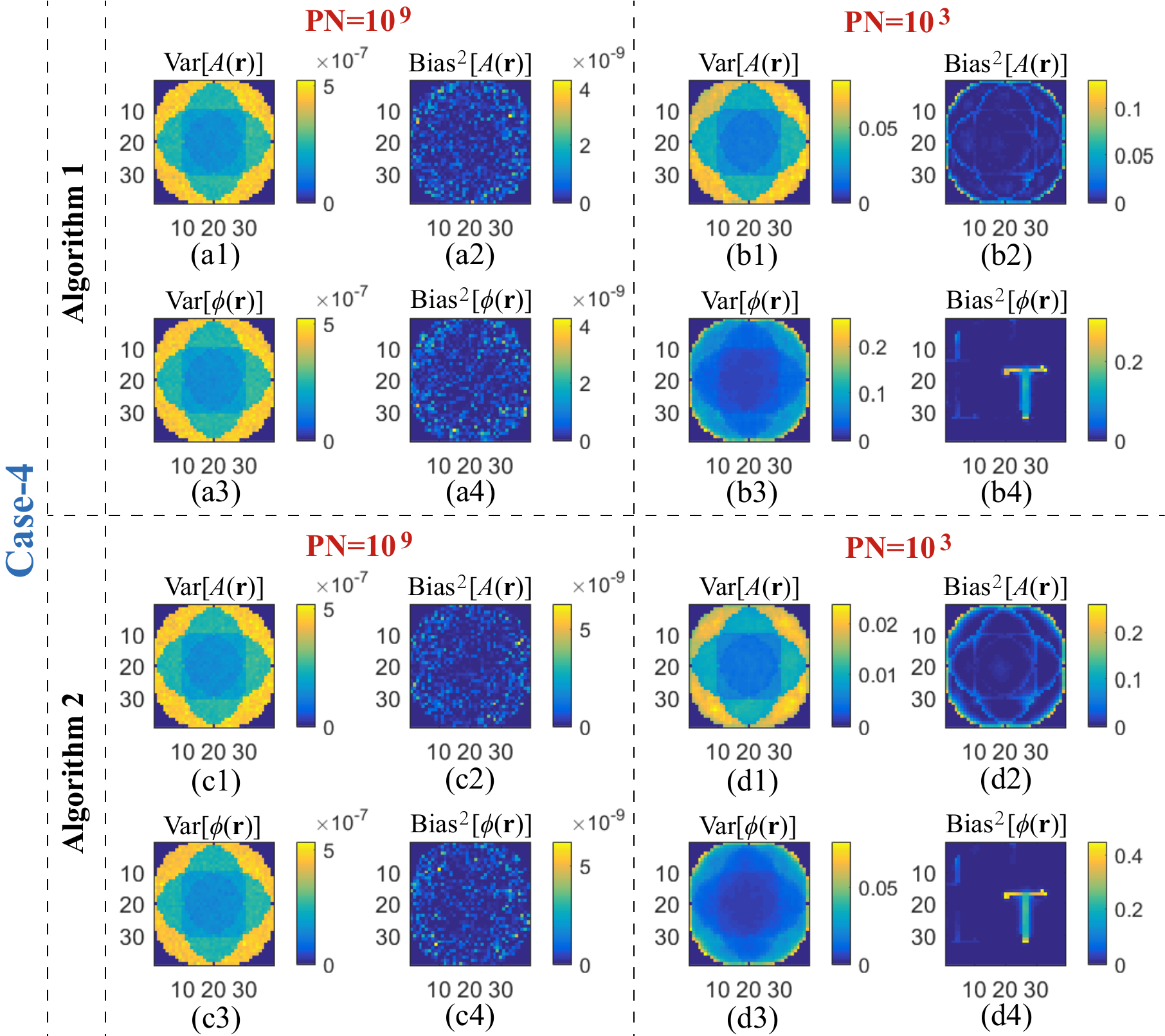}	
	\caption{The Monte Carlo experiment result for Case-4.}\label{Fig.8}
\end{figure}

Next we consider Case-3 where the object has a spatially varying amplitude but the phase is uniform and Case-4, where the amplitude is uniform but the phase has variation. In both cases the probe is a plane wave truncated by a circular aperture. We use the object and probe as in Fig. \ref{Fig.3}a and Fig. \ref{Fig.4}a. The Monte Carlo results obtained with \textbf{Algorithm \ref{alg.1}} and \textbf{Algorithm \ref{alg.2}} for Case-3 are shown in Fig. \ref{Fig.7} and for Case 4 in Fig. \ref{Fig.8}.

When $\text{PN}=10^9$, the variance shown in Fig. \ref{Fig.7} and Fig. \ref{Fig.8} agree with the computed CRLB in Fig. \ref{Fig.3} and Fig. \ref{Fig.4}. To be explicit, the variance of the phase of the object $\phi(\textbf{r})$ is determined by both the object's transmission $A(\textbf{r})$ and the power of the illumination. The part of the object with lower local transmission will have high variance in reconstruction of the phase. On the other hand, the variance of $A(\textbf{r})$ is influenced by the sum of the intensities of the illuminations only. These conclusions are true for both algorithms. Meanwhile, we see that the object itself does not influence the bias of the reconstruction when the photon count is high, which means that both algorithms are unbiased for high photon count.

When the photon number is low, i.e. $\text{PN}=10^3$, the profile of the variance deviates from the computed CRLB which is given in Section 3.2. This statement is true for both \textbf{Algorithm \ref{alg.1}} and \textbf{Algorithm \ref{alg.2}}, and is particularly obvious for $\phi(\textbf{r})$ as shown in Fig. \ref{Fig.7} and Fig. \ref{Fig.8}. We can see that there is trace of the actual $A(\textbf{r})$ in Fig. \ref{Fig.7}b2 and in Fig. \ref{Fig.7}d2, and trace of the actual $\phi(\textbf{r})$ in \ref{Fig.8}b2 and in Fig. \ref{Fig.8}d2, respectively. This trace indicate that, with the current data-set, both two algorithms cannot converge to the CRLB for low photon counts. 

Interestingly, although the object's transmission $A(\textbf{r})$ predominately determines the variance of the object's phase function $\phi(\textbf{r})$, there is no effect of $A$ on the bias of $\phi$ for any value of PN. In the mean time, we see that $\phi$ do not influence the bias of $A$ for any value of PN, as shown in Fig. \ref{Fig.7} and Fig. \ref{Fig.8}. Together with Fig. \ref{Fig.5} and Fig. \ref{Fig.6} in the previous section, we conclude that the profile of the illumination and the object have more influence on the variance of the solutions obtained with \textbf{Algorithm \ref{alg.1}} and \textbf{Algorithm \ref{alg.2}}, more strongly than the amount of bias.

\subsection{The CRLB, variance and bias-variance-ratio in ptychography}
It is seen in Fig. \ref{Fig.5} - Fig. \ref{Fig.8} that the ratio of the bias and the variance, as obtained with both algorithms, tend to increase when the photon count is lower. To further investigate this trend and the property of the two algorithms, we define the bias-variance-ratio (BVR) of the estimator $\hat{O}$ by:
\begin{align}
\text{BVR}\left(\hat{O}\right)\,=\,\frac{\sum_{\textbf{r}}\text{Bias}^{2}\left[\hat{O}(\textbf{r})\right]}{\sum_{\textbf{r}}\text{Var}\left[\hat{O}(\textbf{r})\right]}.\label{BVR}
\end{align}

\begin{figure}[htp!]
	\centering\includegraphics[width=0.9\textwidth]{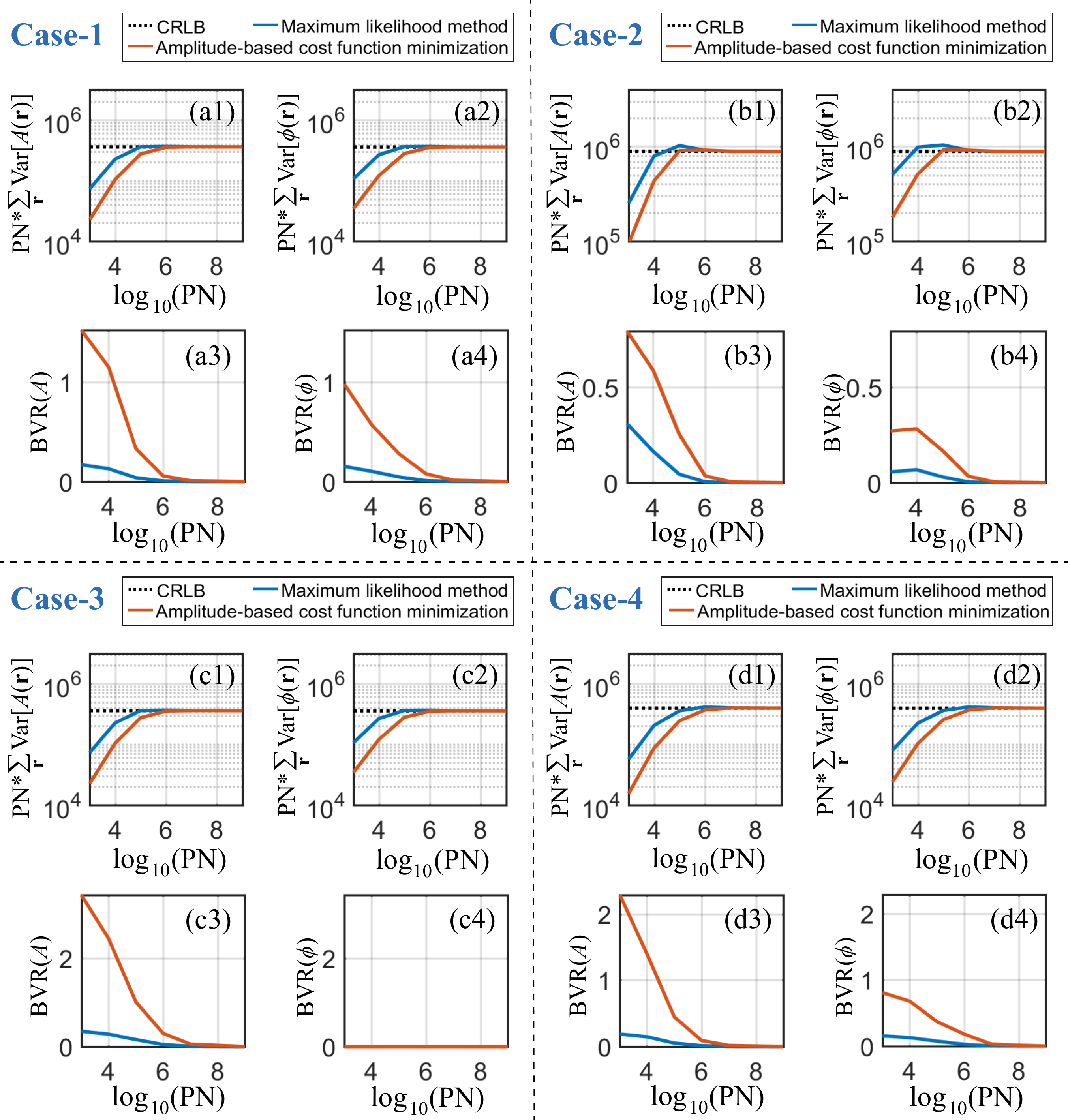}	\caption{The CRLB, variance and bias-variance-ratio of two algorithms for various of values of $\text{PN}$.}\label{Fig.9}
\end{figure}

In Fig. \ref{Fig.9} we show the BVR of \textbf{Algorithm \ref{alg.1}} and \textbf{Algorithm \ref{alg.2}} for various photon counts and for Case-1 to Case-4. 
The overall CRLB and variance of $A(\textbf{r})$ and $\phi(\textbf{r})$ obtained from both algorithms are also shown. We see that the overall variance of both algorithms are the same as the computed CRLB asymptotically when the photon number is high. For lower photon counts, the variance become lower than the CRLB, meanwhile the BVR of both algorithms increase. For our current configuration, this threshold is at $\text{PN}=10^6$. When $\text{PN}<10^6$, the variance of \textbf{Algorithm \ref{alg.1}} is higher than \textbf{Algorithm \ref{alg.2}} for all Case-1 to Case-4. On the other hand, the BVR of \textbf{Algorithm \ref{alg.1}} is higher than \textbf{Algorithm \ref{alg.2}}, which indicates that the \textbf{Algorithm \ref{alg.1}} generally has lower bias than \textbf{Algorithm \ref{alg.2}}. 

\subsection{Discussion}
It is seen in the Monte Carlo results that, for low photon counts, the variance with both \textbf{Algorithm \ref{alg.1}} and \textbf{Algorithm \ref{alg.2}} are lower than the computed CRLB. This observation indicates that, with the current data-set, the two estimators are unbiased for high photon counts but cannot convergen to the CRLB when the photon number is low. 

One may argue that the variances shown in Fig. \ref{Fig.9} are lower than the CRLB when PN$<10^{6}$ because the current data-set is insufficient\cite{Kay2009}. In particular, if sufficient amount of data is given, the maximum likelihood estimator should be asymptotically unbiased and achieves the CRLB if sufficient amount of data is given, as shown in Eq. (\ref{eq.MLE}). Indeed, we see in the simulation that Eq. (\ref{eq.MLE}) holds when PN$>10^{6}$, which indicates that the current data-set is already sufficient when PN$>10^{6}$. However, for low photon counts, the current data-set is insufficient for the maximum likelihood estimator to converge to the CRLB. 

To explain this fact, we first investigate the signal-to-noise ratio (SNR) of each $m$th ptychographyic measurement with Poisson noise:
\begin{align}
\text{SNR}_{P,m}(\bm{\upxi})\,=\,\sqrt{n_{m}(\bm{\upxi})}.\label{SNR}
\end{align}
For typical far-field diffraction patterns the intensities are not uniform. Hence, the SNR should be a function of $\bm{\upxi}$ and the value of SNR should vary per pixel on the detector. Nevertheless, we can still see that the SNR will in general decrease when the number of photon detected is decreased. Therefore, for Poisson noise, one can extract less and less information about the actual signal when the photon counts is decreasing. 

Moreover, we note that the measurement $n_{m}(\bm{\upxi})$ is discontinuous and contains nature numbers only, which is associated with the particle nature of light or the quantization error that occurs in the detector. This discontinuity has more disruptive effect on the measurement for the case of low photon counts than the case of high photon number. Taking an extreme example, suppose only one photon is detected, this photon will most likely appears at $\bm{\upxi}=0$. Therefore, almost all of the spatial information about the object are lost in the measurement, and hence it is more difficult for estimators to converge to the CRLB.

\begin{figure}[htp!]
	\centering\includegraphics[width=0.88\textwidth]{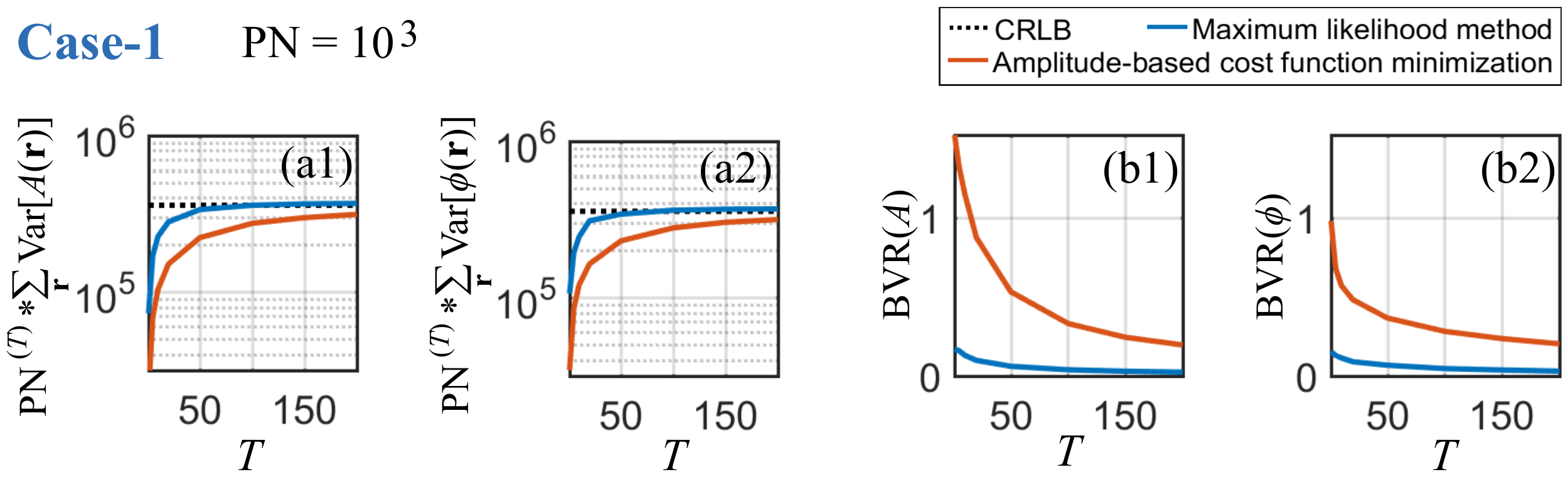}	\caption{The CRLB, variance and bias-variance-ratio of two algorithms for various of number of measurements $T$. This plot is for Case-1 and for PN$=10^{3}$.}\label{Fig.10}
\end{figure}

If we want to increase the size of data-set while keeping the current characteristic simulation parameters, one way is to take multiple measurements for each $m$th probe's position. Suppose for each probe's position we take $T$ measurements, denoted by: $n_{m,t}(\bm{\upxi})$, where $t=1,2,\cdots,T$. A straightforward way to process the data is simply to compute the mean of the measurements:
\begin{align}
n^{(T)}_{m}(\bm{\upxi})\,=\,\frac{\sum_{t}n_{m,t}(\bm{\upxi})}{T}.\label{AveragedData}
\end{align}
It has been shown that, when $T$ is large enough, Eq. (\ref{AveragedData}) is a sufficient statistic for Poisson distribution. That is, $n^{(T)}_{m}$ carries all the information as in the data-set: $n_{m,t}$, $t=1,2,\cdots,T$. In Fig. \ref{Fig.10} the Monte Carlo result with data-set $n_{m,t}(\bm{\upxi})$ is shown. To give an example, we study Case-1 for low photon counts, i.e. PN=$10^{3}$. We note that, by summing over all $T$ measurements, the total photon number $\text{PN}^{(T)}$ counting in the probe is now given by:
\begin{align}
\text{PN}^{(T)}\,=\,\dfrac{\sum_{\textbf{r}}\left|P(\textbf{r})\right|^{2}}{\hbar\omega}*T\,=\,\text{PN}*T,\label{PN^T}
\end{align}
and the CRLB is proportional to the reciprocal of $\text{PN}^{(T)}$ according to Eq. (\ref{eq.21}).

Fig. \ref{Fig.10}a shows the computed CRLB and the variance of reconstruction for various of number of measurements $T$. We see that, for both two algorithms, the variances approach the CRLB as the number of measurements is increasing. In particular, the variance of \textbf{Algorithm \ref{alg.1}} have reached the CRLB when $T$ is up to 200. Meanwhile, it is seen in Fig. \ref{Fig.10}b that the bias of \textbf{Algorithm \ref{alg.1}} is considerably small comparing to the variance when $T>200$. Therefore, we confirm that, for low photon counts, \textbf{Algorithm \ref{alg.1}} can be asymptotically unbiased and converge to the CRLB by increasing the number of measurements. We see in Fig. \ref{Fig.10} that this conclusion is true for \textbf{Algorithm \ref{alg.2}} also. However, the speed of this convergence for \textbf{Algorithm \ref{alg.2}} is slower than for \textbf{Algorithm \ref{alg.1}}.

\section{Conclusion}
In the first part of this paper we have studied the influence of Poisson noise on ptychography by analyzing the CRLB. The CRLB was theoretically derived and numerically computed from the Fisher matrix for 4 different cases. It was found that if the estimator is unbiased, the minimum variance in the presence of Poisson noise is mostly determined by both the illumination's local dose distribution and the object's local transmission. The calculations of the CRLB suggest that the minimum variance is inversely proportional to the number of photons in the illumination beam. The computations of the CRLB using the Fisher matrix were validated with Monte Carlo analysis. It was confirmed that the local illumination power has a strong effect on the variance of the reconstruction of both object's transmission and phase function. Meanwhile the object's actual local transmission strongly influences the reconstruction of the object's phase. 

In the second part of this work, the statistical properties of the maximum likelihood method and the amplitude-based cost function minimization algorithm are studied. Both algorithms were applied in the Monte Carlo simulations, using a conjugate gradient based implementation. It was shown that both approaches are asymptotically unbiased with variances that are slightly larger than the CRLB when the photon counts is high. For the case of lower photon number, the Monte Carlo analysis showed that both method require more measurement to converge to the CRLB and to be estimators. While increasing the number of data, it was shown that the maximum likelihood method converges to the CRLB faster than the amplitude-based cost function minimization algorithm.

Our result can help to understand the defects that occur in the ptychograghy reconstruction from noisy data. Our conclusions suggest that more illumination power should be given to the part of object which is of most interest. As next steps of research, the performance of other ptychographic de-noising algorithm \cite{Wen2012,Horstmeyer2015,Maiden2017,Konijnenberg2018a,Pham2019} deserve further investigation. Investigating the CRLB and the statistic properties of the two algorithms for Gaussian noise and the mixed Poisson-Gaussian noise is also an interesting topic for further research.

\appendix 	
\section*{Appendix:}
The detail of \textbf{Algorithm \ref{alg.1}} is described in the pseudo-code.
\begin{algorithm}
	\caption{Maximum likelihood method with Poisson noise}
	\label{alg.1}
	\begin{algorithmic}[1]
		\STATE $k_{\text{max}}\gets 10^{3}$, $\delta_{\mathcal{L}}\gets 10^{-20}$, $\gamma\gets 10^{-5}$, $A_{1} \gets A_{\text{o}}$, $\phi_{1} \gets \phi_{\text{o}}$, $k \gets 1$.
		\REPEAT
		\STATE compute the steepest descent gradient of $A$ and $\phi$ using Eq. (\ref{eq.12}):\\
		$g_{A,k} \gets \sum_{{m}}-\Re\left\lbrace P^{*}_{m}e^{-\text{i}\phi_{k}} \mathcal{F}^{-1}\left[\left(\frac{n_{m}}{N_{m}+\gamma}-1\right)\mathcal{F}\left(P_{m}O_{k}\right)\right]\right\rbrace$,\\
		$g_{\phi,k} \gets \sum_{{m}}-\Im\left\lbrace P^{*}_{m}A_{k} e^{-\text{i}\phi_{k}} \mathcal{F}^{-1}\left[\left(\frac{n_{m}}{N_{m}+\gamma}-1\right)\mathcal{F}\left(P_{m}O_{k}\right)\right]\right\rbrace$.
		\IF { $k=1$ }
		\STATE $\Delta_{A,k} \gets g_{A,k}$,\quad$\Delta_{\phi,k} \gets g_{\phi,k}$.
		\ELSE 
		\STATE use the formula of Polak–Ribi\`{e}re:\\ $\beta^{\text{PR}}_{A,k} \gets \dfrac{\langle\left(g_{A,k}-g_{A,k-1}\right) |g_{A,k}\rangle}{\left\|g_{A,k-1}\right\|_{2}^{2}}$,\quad
		$\beta^{\text{PR}}_{\phi,k} \gets \dfrac{\langle\left(g_{\phi,k}-g_{\phi,k-1}\right) |g_{\phi,k}\rangle}{\left\|g_{\phi,k-1}\right\|_{2}^{2}}$,
		\STATE $\beta_{A,k} \gets \max\left(\beta^{\text{PR}}_{A,k},0\right)$,\quad$\beta_{\phi,k} \gets \max\left(\beta^{\text{PR}}_{\phi,k},0\right)$,
		\STATE compute the conjugate direction: \\
		$\Delta_{A,k} \gets g_{A,k}+\beta_{A,k}\Delta_{A,k-1}$,\quad$\Delta_{\phi,k} \gets g_{\phi,k}+\beta_{\phi,k}\Delta_{\phi,k-1}$.
		\ENDIF.
		\STATE optimize the update step size: \\
		$\alpha_{A,k} \gets \arg\min\limits_{\alpha_{A}}\mathcal{L}_{P}\left(A_{k}+\alpha_{A}\Delta_{A,k}\right)$,\quad$\alpha_{\phi,k} \gets \arg\min\limits_{\alpha_{\phi}}\mathcal{L}_{P}\left(\phi_{k}+\alpha_{\phi}\Delta_{\phi,k}\right)$.
		\STATE update the object function: \\
		$A_{k+1} \gets A_{k}+\alpha_{A,k}\Delta_{A,k}$,\quad$\phi_{k+1} \gets \phi_{k}+\alpha_{\phi,k}\Delta_{\phi,k}$.
		\IF { $k=11$ }
		\STATE $\gamma \gets 10^{-20}$,
		\ENDIF.
		\STATE $k \gets k+1$.
		\UNTIL{ $k=k_{\text{max}}$ or $\left|\mathcal{L}_{P,k}-\mathcal{L}_{P,k-1}\right|\leq\delta_{\mathcal{L}}$}.
	\end{algorithmic}
\end{algorithm}
Unlike Eq. (\ref{eq.12}), the update step size $\alpha$ is not a constant anymore in \textbf{Algorithm \ref{alg.1}}. Instead, an optimal $\alpha$ for every iteration $k$ is obtained in the manner described in \cite{Coene1996}: (1) Based on the computed $k$th local gradient, calculate the value of the likelihood function $\mathcal{L}_P$ for at least three different values of $\alpha$, e.g. [0.01,0.5,1]. (2) Approximate $\mathcal{L}_{P}$ by a quadratic function of $\alpha$. To do this we apply the 'polyfit' routine in MATLAB. (3) Choose the value for $\alpha$ for which the quadratic function is minimum. The parameter $\beta_{k}$ is chosen such that the update direction of the object function is conjugate between two subsequent iterations, for which many proposals exist \cite{Fletcher1988}. Based on the formula of Polak–Ribi\`{e}re\cite{Shewchuk1994AnIT}, we choose $\beta_{k}=\max\left(\beta^{\text{PR}}_{k},0\right)$, where $\beta^{\text{PR}}_{k}$ is given by:
\begin{align}
\beta^{\text{PR}}_{k}\,=\,\frac{\langle\left(g_{k}-g_{k-1}\right) |g_{k}\rangle}{\left\|g_{k-1}\right\|_{2}^{2}},\label{PR}
\end{align}
where $g_{k}$ is the gradient of $\mathcal{L}_{P}$ with respect to $O(\textbf{r})$ in the $k$th iteration. When the calculated $\beta^{\text{PR}}_{k}$ have negative value, $\beta_{k}$ resets the search direction from the conjugate gradient back to the local decent gradient direction, i.e. $\Delta_{k} \gets g_{k}$.

In order to prevent that the algorithm terminates in a local minimum, the initial guess of the object is selected to be the actual object $A_{\text{o}}(\textbf{r})$ and $\phi_{\text{o}}(\textbf{r})$.  The denominator $N_{m}$ in Eq. (\ref{eq.12}) is a function of $\bm{\upxi}$, and may be close to zero for some $\bm{\upxi}$. Hence the maximum likelihood method can be unstable. To avoid the instability, a regularization parameter $\gamma$ is introduced in \textbf{Algorithm \ref{alg.1}}, of which the value can be determined in practice depending on the noise level. Throughout this paper, we let $\gamma$ be $10^{-5}$ (note that $N_{m}$ is non-negative integer) for the first 10 iterations, then reset $\gamma$ to $10^{-20}$ after the 10th iteration. \textbf{Algorithm \ref{alg.1}} terminates when the change of the likelihood function between two subsequent iterations is smaller than a threshold $\delta_{\mathcal{L}}$, or when the number of iteration reaches a maximum $k_{\text{max}}$.
\begin{algorithm}
	\caption{Amplitude-based cost function minimization approach}
	\label{alg.2}
	\begin{algorithmic}[1]
		\STATE $k_{\text{max}}\gets 10^{3}$, $\delta_{\mathcal{E}}\gets 10^{-20}$, $\gamma\gets 10^{-3}$, $A_{1} \gets A_{\text{o}}$, $\phi_{1} \gets \phi_{\text{o}}$, $k \gets 1$, 
		\REPEAT
		\STATE compute the steepest descent gradient of $A$ and $\phi$:\\
		$g_{A,k} \gets \sum_{{m}}-\Re\left\lbrace P^{*}_{m}e^{-\text{i}\phi_{k}} \mathcal{F}^{-1}\left[\left(\frac{\sqrt{n_{m}}}{\sqrt{N_{m}+\gamma}}-1\right)\mathcal{F}\left(P_{m}O_{k}\right)\right]\right\rbrace$,\\
		$g_{\phi,k} \gets \sum_{{m}}-\Im\left\lbrace P^{*}_{m}A_{k} e^{-\text{i}\phi_{k}} \mathcal{F}^{-1}\left[\left(\frac{\sqrt{n_{m}}}{\sqrt{N_{m}+\gamma}}-1\right)\mathcal{F}\left(P_{m}O_{k}\right)\right]\right\rbrace$.
		\STATE follow 4th-10th steps of \textbf{Algorithm \ref{alg.1}}.
		\STATE optimize the update step size: \\
		$\alpha_{A,k} \gets \arg\min\limits_{\alpha_{A}}\mathcal{E}\left(A_{k}+\alpha_{A}\Delta_{A,k}\right)$,\quad$\alpha_{\phi,k} \gets \arg\min\limits_{\alpha_{\phi}}\mathcal{E}\left(\phi_{k}+\alpha_{\phi}\Delta_{\phi,k}\right)$.
		\STATE follow 12th-16th steps of \textbf{Algorithm \ref{alg.1}}.
		\UNTIL{ $k=k_{\text{max}}$ or $\left|\mathcal{E}_{k}-\mathcal{E}_{k-1}\right|\leq\delta_{\mathcal{E}}$}.
	\end{algorithmic}
\end{algorithm}

For comparison, the performance of another popular method, namely the amplitude-based cost function minimization approach \cite{Guizar-Sicairos2008}, is investigated in the Monte Carlo experiment. The approach is described  in \textbf{Algorithm \ref{alg.2}}, in which the search of the optimal step size $\alpha_{k}$ and the method of conjugate gradient are added too. Similar to \textbf{Algorithm \ref{alg.1}}, \textbf{Algorithm \ref{alg.2}} stops when the change of the cost function between two subsequent iterations is smaller than a threshold $\delta_{\mathcal{E}}$, or when the number of iteration reaches a maximum $k_{\text{max}}$.

\section*{Funding}
H2020 Marie Sk$\nmid$odowska-Curie Actions (675745).

\section*{Acknowledgments}	
X. Wei thanks Z. Xi for fruitful discussions.

\section*{Disclosures}
The authors declare no conflicts of interest.

\bibliography{sample}

\begin{thebibliography}{10}
\newcommand{\enquote}[1]{``#1''}

\bibitem{Hoppe1969}
W.~Hoppe, \enquote{Beugung im inhomogenen primärstrahlwellenfeld. i. prinzip
  einer phasenmessung von elektronenbeungungsinterferenzen,}
  {\protect\JournalTitle{Acta Crystallographica Section A}} \textbf{25},
  495--501 (1969).

\bibitem{Rodenburg1992a}
J.~M. Rodenburg and R.~H.~T. Bates, \enquote{The theory of super-resolution
  electron microscopy via wigner-distribution deconvolution,}
  {\protect\JournalTitle{Philosophical Transactions of the Royal Society of
  London. Series A: Physical and Engineering Sciences}} \textbf{339}, 521--553
  (1992).

\bibitem{Chapman1996}
H.~N. Chapman, \enquote{Phase-retrieval x-ray microscopy by wigner-distribution
  deconvolution,} {\protect\JournalTitle{Ultramicroscopy}} \textbf{66},
  153--172 (1996).

\bibitem{Faulkner2004}
H.~M.~L. Faulkner and J.~M. Rodenburg, \enquote{Movable aperture lensless
  transmission microscopy: A novel phase retrieval algorithm,}
  {\protect\JournalTitle{Physical Review Letters}} \textbf{93}, 023903 (2004).

\bibitem{Rodenburg2004}
J.~M. Rodenburg and H.~M.~L. Faulkner, \enquote{A phase retrieval algorithm for
  shifting illumination,} {\protect\JournalTitle{Applied Physics Letters}}
  \textbf{85}, 4795--4797 (2004).

\bibitem{Guizar-Sicairos2008}
M.~Guizar-Sicairos and J.~R. Fienup, \enquote{Phase retrieval with transverse
  translation diversity: a nonlinear optimization approach,}
  {\protect\JournalTitle{Optics Express}} \textbf{16}, 7264--7278 (2008).

\bibitem{Silva2015}
J.~C. da~Silva and A.~Menzel, \enquote{Elementary signals in ptychography,}
  {\protect\JournalTitle{Optics Express}} \textbf{23}, 33812--33821 (2015).

\bibitem{Seaberg2014}
M.~D. Seaberg, B.~Zhang, D.~F. Gardner, E.~R. Shanblatt, M.~M. Murnane, H.~C.
  Kapteyn, and D.~E. Adams, \enquote{Tabletop nanometer extreme ultraviolet
  imaging in an extended reflection mode using coherent fresnel ptychography,}
  {\protect\JournalTitle{Optica}} \textbf{1}, 39--44 (2014).

\bibitem{Odstrcil2015}
M.~Odstrcil, J.~Bussmann, D.~Rudolf, R.~Bresenitz, J.~Miao, W.~S. Brocklesby,
  and L.~Juschkin, \enquote{Ptychographic imaging with a compact
  gas{\textendash}discharge plasma extreme ultraviolet light source,}
  {\protect\JournalTitle{Optics Letters}} \textbf{40}, 5574--5577 (2015).

\bibitem{Rodenburg2007}
J.~M. Rodenburg, A.~C. Hurst, A.~G. Cullis, B.~R. Dobson, F.~Pfeiffer, O.~Bunk,
  C.~David, K.~Jefimovs, and I.~Johnson, \enquote{Hard-x-ray lensless imaging
  of extended objects,} {\protect\JournalTitle{Physical Review Letters}}
  \textbf{98}, 034801 (2007).

\bibitem{Thibault2008}
P.~Thibault, M.~Dierolf, A.~Menzel, O.~Bunk, C.~David, and F.~Pfeiffer,
  \enquote{High-resolution scanning x-ray diffraction microscopy,}
  {\protect\JournalTitle{Science}} \textbf{321}, 379--382 (2008).

\bibitem{Chapman2010}
H.~N. Chapman and K.~A. Nugent, \enquote{Coherent lensless x-ray imaging,}
  {\protect\JournalTitle{Nature Photonics}} \textbf{4}, 833--839 (2010).

\bibitem{Pfeiffer2017}
F.~Pfeiffer, \enquote{X-ray ptychography,} {\protect\JournalTitle{Nature
  Photonics}} \textbf{12}, 9--17 (2017).

\bibitem{Thibault2009}
P.~Thibault, M.~Dierolf, O.~Bunk, A.~Menzel, and F.~Pfeiffer, \enquote{Probe
  retrieval in ptychographic coherent diffractive imaging,}
  {\protect\JournalTitle{Ultramicroscopy}} \textbf{109}, 338--343 (2009).

\bibitem{Maiden2009}
A.~M. Maiden and J.~M. Rodenburg, \enquote{An improved ptychographical phase
  retrieval algorithm for diffractive imaging,}
  {\protect\JournalTitle{Ultramicroscopy}} \textbf{109}, 1256--1262 (2009).

\bibitem{Holler2017}
M.~Holler, M.~Guizar-Sicairos, E.~H.~R. Tsai, R.~Dinapoli, E.~Müller, O.~Bunk,
  J.~Raabe, and G.~Aeppli, \enquote{High-resolution non-destructive
  three-dimensional imaging of integrated circuits,}
  {\protect\JournalTitle{Nature}} \textbf{543}, 402--406 (2017).

\bibitem{Gardner2017}
D.~F. Gardner, M.~Tanksalvala, E.~R. Shanblatt, X.~Zhang, B.~R. Galloway, C.~L.
  Porter, R.~K. Jr, C.~Bevis, D.~E. Adams, H.~C. Kapteyn, M.~M. Murnane, and
  G.~F. Mancini, \enquote{Subwavelength coherent imaging of periodic samples
  using a 13.5{\hspace{0.25em}}nm tabletop high-harmonic light source,}
  {\protect\JournalTitle{Nature Photonics}} \textbf{11}, 259--263 (2017).

\bibitem{Jiang2018}
Y.~Jiang, Z.~Chen, Y.~Han, P.~Deb, H.~Gao, S.~Xie, P.~Purohit, M.~W. Tate,
  J.~Park, S.~M. Gruner, V.~Elser, and D.~A. Muller, \enquote{Electron
  ptychography of 2d materials to deep sub-{\aa}ngström resolution,}
  {\protect\JournalTitle{Nature}} \textbf{559}, 343--349 (2018).

\bibitem{Maiden2017}
A.~Maiden, D.~Johnson, and P.~Li, \enquote{Further improvements to the
  ptychographical iterative engine,} {\protect\JournalTitle{Optica}}
  \textbf{4}, 736--745 (2017).

\bibitem{Zheng2013}
G.~Zheng, R.~Horstmeyer, and C.~Yang, \enquote{Wide-field, high-resolution
  fourier ptychographic microscopy,} {\protect\JournalTitle{Nature Photonics}}
  \textbf{7}, 739--745 (2013).

\bibitem{Yeh2015}
L.-H. Yeh, J.~Dong, J.~Zhong, L.~Tian, M.~Chen, G.~Tang, M.~Soltanolkotabi, and
  L.~Waller, \enquote{Experimental robustness of fourier ptychography phase
  retrieval algorithms,} {\protect\JournalTitle{Optics Express}} \textbf{23},
  33214--33240 (2015).

\bibitem{Zhang2017}
Y.~Zhang, P.~Song, and Q.~Dai, \enquote{Fourier ptychographic microscopy using
  a generalized anscombe transform approximation of the mixed poisson-gaussian
  likelihood,} {\protect\JournalTitle{Optics Express}} \textbf{25}, 168--179
  (2017).

\bibitem{Thibault13}
P.~Thibault and A.~Menzel, \enquote{Reconstructing state mixtures from
  diffraction measurements,} {\protect\JournalTitle{Nature}} \textbf{494},
  68--71 (2013).

\bibitem{Burdet2015}
N.~Burdet, X.~Shi, D.~Parks, J.~N. Clark, X.~Huang, S.~D. Kevan, and I.~K.
  Robinson, \enquote{Evaluation of partial coherence correction in x-ray
  ptychography,} {\protect\JournalTitle{Optics Express}} \textbf{23},
  5452--5467 (2015).

\bibitem{Zhong2016}
J.~Zhong, L.~Tian, P.~Varma, and L.~Waller, \enquote{Nonlinear optimization
  algorithm for partially coherent phase retrieval and source recovery,}
  {\protect\JournalTitle{{IEEE} Transactions on Computational Imaging}}
  \textbf{2}, 310--322 (2016).

\bibitem{Batey2014}
D.~J. Batey, D.~Claus, and J.~M. Rodenburg, \enquote{Information multiplexing
  in ptychography,} {\protect\JournalTitle{Ultramicroscopy}} \textbf{138},
  13--21 (2014).

\bibitem{Wei2019}
X.~Wei and P.~Urbach, \enquote{Ptychography with multiple wavelength
  illumination,} {\protect\JournalTitle{Optics Express}} \textbf{27},
  36767--36789 (2019).

\bibitem{Maiden2012}
A.~M. Maiden, M.~J. Humphry, and J.~M. Rodenburg, \enquote{Ptychographic
  transmission microscopy in three dimensions using a multi-slice approach,}
  {\protect\JournalTitle{Journal of the Optical Society of America A}}
  \textbf{29}, 1606--1614 (2012).

\bibitem{Gilles2018}
M.~A. Gilles, Y.~S.~G. Nashed, M.~Du, C.~Jacobsen, and S.~M. Wild, \enquote{3d
  x-ray imaging of continuous objects beyond the depth of focus limit,}
  {\protect\JournalTitle{Optica}} \textbf{5}, 1078--1086 (2018).

\bibitem{Kahnt2019}
M.~Kahnt, J.~Becher, D.~Brückner, Y.~Fam, T.~Sheppard, T.~Weissenberger,
  F.~Wittwer, J.-D. Grunwaldt, W.~Schwieger, and C.~G. Schroer,
  \enquote{Coupled ptychography and tomography algorithm improves
  reconstruction of experimental data,} {\protect\JournalTitle{Optica}}
  \textbf{6}, 1282--1289 (2019).

\bibitem{Pelz2014}
P.~M. Pelz, M.~Guizar-Sicairos, P.~Thibault, I.~Johnson, M.~Holler, and
  A.~Menzel, \enquote{On-the-fly scans for x-ray ptychography,}
  {\protect\JournalTitle{Applied Physics Letters}} \textbf{105}, 251101 (2014).

\bibitem{Deng2015}
J.~Deng, Y.~S.~G. Nashed, S.~Chen, N.~W. Phillips, T.~Peterka, R.~Ross,
  S.~Vogt, C.~Jacobsen, and D.~J. Vine, \enquote{Continuous motion scan
  ptychography: characterization for increased speed in coherent x-ray
  imaging,} {\protect\JournalTitle{Optics Express}} \textbf{23}, 5438–5451
  (2015).

\bibitem{Flaes2018}
D.~E.~B. Flaes and S.~Witte, \enquote{Interference probe ptychography for
  computational amplitude and phase microscopy,} {\protect\JournalTitle{Optics
  Express}} \textbf{26}, 31372--31390 (2018).

\bibitem{Elser2003}
V.~Elser, \enquote{Phase retrieval by iterated projections,}
  {\protect\JournalTitle{Journal of the Optical Society of America A}}
  \textbf{20}, 40--55 (2003).

\bibitem{Wen2012}
Z.~Wen, C.~Yang, X.~Liu, and S.~Marchesini, \enquote{Alternating direction
  methods for classical and ptychographic phase retrieval,}
  {\protect\JournalTitle{Inverse Problems}} \textbf{28}, 115010 (2012).

\bibitem{Marchesini2013}
S.~Marchesini, A.~Schirotzek, C.~Yang, H.~tieng Wu, and F.~Maia,
  \enquote{Augmented projections for ptychographic imaging,}
  {\protect\JournalTitle{Inverse Problems}} \textbf{29}, 115009 (2013).

\bibitem{Horstmeyer2015}
R.~Horstmeyer, R.~Y. Chen, X.~Ou, B.~Ames, J.~A. Tropp, and C.~Yang,
  \enquote{Solving ptychography with a convex relaxation,}
  {\protect\JournalTitle{New Journal of Physics}} \textbf{17}, 053044 (2015).

\bibitem{Pham2019}
M.~Pham, A.~Rana, J.~Miao, and S.~Osher, \enquote{Semi-implicit relaxed
  douglas-rachford algorithm ({sDR}) for ptychography,}
  {\protect\JournalTitle{Optics Express}} \textbf{27}, 31246--31260 (2019).

\bibitem{Fannjiang2019}
A.~Fannjiang and P.~Chen, \enquote{Blind ptychography: Uniqueness and
  ambiguities,} {\protect\JournalTitle{arXiv: 1806.02674v3}}  (2018).

\bibitem{Thibault2012}
P.~Thibault and M.~Guizar-Sicairos, \enquote{Maximum-likelihood refinement for
  coherent diffractive imaging,} {\protect\JournalTitle{New Journal of
  Physics}} \textbf{14}, 063004 (2012).

\bibitem{Godard2012}
P.~Godard, M.~Allain, V.~Chamard, and J.~Rodenburg, \enquote{Noise models for
  low counting rate coherent diffraction imaging,}
  {\protect\JournalTitle{Optics Express}} \textbf{20}, 25914--25934 (2012).

\bibitem{Chang2019}
H.~Chang, P.~Enfedaque, J.~Zhang, J.~Reinhardt, B.~Enders, Y.-S. Yu,
  D.~Shapiro, C.~G. Schroer, T.~Zeng, and S.~Marchesini, \enquote{Advanced
  denoising for x-ray ptychography,} {\protect\JournalTitle{Optics Express}}
  \textbf{27}, 10395--10418 (2019).

\bibitem{Suzuki2015}
A.~Suzuki and Y.~Takahashi, \enquote{Dark-field x-ray ptychography,}
  {\protect\JournalTitle{Optics Express}} \textbf{23}, 16429--16438 (2015).

\bibitem{Stockmar2013}
M.~Stockmar, P.~Cloetens, I.~Zanette, B.~Enders, M.~Dierolf, F.~Pfeiffer, and
  P.~Thibault, \enquote{Near-field ptychography: phase retrieval for inline
  holography using a structured illumination,}
  {\protect\JournalTitle{Scientific Reports}} \textbf{3}, 1927 (2013).

\bibitem{Zuo2016}
C.~Zuo, J.~Sun, and Q.~Chen, \enquote{Adaptive step-size strategy for
  noise-robust fourier ptychographic microscopy,} {\protect\JournalTitle{Optics
  Express}} \textbf{24}, 20724--20744 (2016).

\bibitem{Odstrcil2018}
M.~Odstr{\v{c}}il, A.~Menzel, and M.~Guizar-Sicairos, \enquote{Iterative
  least-squares solver for generalized maximum-likelihood ptychography,}
  {\protect\JournalTitle{Optics Express}} \textbf{26}, 3108--3123 (2018).

\bibitem{Konijnenberg2018a}
A.~P. Konijnenberg, W.~M.~J. Coene, and H.~P. Urbach,
  \enquote{Model-independent noise-robust extension of ptychography,}
  {\protect\JournalTitle{Optics Express}} \textbf{26}, 5857--5874 (2018).

\bibitem{Bartlett1936}
M.~S. Bartlett, \enquote{The square root transformation in analysis of
  variance,} {\protect\JournalTitle{Supplement to the Journal of the Royal
  Statistical Society}} \textbf{3}, 68--78 (1936).

\bibitem{Anscombe1948}
F.~J. Anscombe, \enquote{The transformation of poisson, binomial and
  negative-binomial data,} {\protect\JournalTitle{Biometrika}} \textbf{35},
  246--254 (1948).

\bibitem{Kay2009}
S.~M. Kay, \emph{Fundamentals Of Statistical Signal Processing, Volume 1:
  Estimation Theory} (Pearson, 2009).

\bibitem{Cederquist1987}
J.~N. Cederquist and C.~C. Wackerman, \enquote{Phase-retrieval error: a lower
  bound,} {\protect\JournalTitle{Journal of the Optical Society of America A}}
  \textbf{4}, 1788--1792 (1987).

\bibitem{Fienup1993}
J.~R. Fienup, J.~C. Marron, T.~J. Schulz, and J.~H. Seldin, \enquote{Hubble
  space telescope characterized by using phase-retrieval algorithms,}
  {\protect\JournalTitle{Applied Optics}} \textbf{32}, 1747--1767 (1993).

\bibitem{goodman2005}
J.~Goodman, \emph{Introduction to Fourier Optics}, McGraw-Hill physical and
  quantum electronics series (W. H. Freeman, 2005).

\bibitem{Murray1982}
W.~Murray, M.~H. Wright, and P.~E. Gill, \emph{Practical Optimization} (Emerald
  Publishing Limited, 1982).

\bibitem{Bouchet2020}
D.~Bouchet, R.~Carminati, and A.~P. Mosk, \enquote{Influence of the local
  scattering environment on the localization precision of single particles,}
  {\protect\JournalTitle{Physical Review Letters}} \textbf{124} (2020).

\bibitem{Bunk2008}
O.~Bunk, M.~Dierolf, S.~Kynde, I.~Johnson, O.~Marti, and F.~Pfeiffer,
  \enquote{Influence of the overlap parameter on the convergence of the
  ptychographical iterative engine,} {\protect\JournalTitle{Ultramicroscopy}}
  \textbf{108}, 481--487 (2008).

\bibitem{Tripathi2014}
A.~Tripathi, I.~McNulty, and O.~G. Shpyrko, \enquote{Ptychographic overlap
  constraint errors and the limits of their numerical recovery using conjugate
  gradient descent methods,} {\protect\JournalTitle{Optics Express}}
  \textbf{22}, 1452--1466 (2014).

\bibitem{Coene1996}
W.~Coene, A.~Thust, M.~O. de~Beeck, and D.~V. Dyck, \enquote{Maximum-likelihood
  method for focus-variation image reconstruction in high resolution
  transmission electron microscopy,} {\protect\JournalTitle{Ultramicroscopy}}
  \textbf{64}, 109--135 (1996).

\bibitem{Fletcher1988}
R.~Fletcher, \emph{Practical Methods of Optimization, 2nd Edition} (Wiley,
  1988).

\bibitem{Shewchuk1994AnIT}
J.~R. Shewchuk, \enquote{An introduction to the conjugate gradient method
  without the agonizing pain,} Tech. rep., Carnegie Mellon University (1994).

\end{thebibliography}
	
\end{document}